\documentclass{aa}  
\usepackage{graphicx}
\usepackage{physics,amsmath}
\usepackage{txfonts}

\usepackage{natbib}
\bibpunct{(}{)}{;}{a}{}{,} 
\usepackage[colorlinks=true, allcolors = blue]{hyperref}

\hypersetup{
  colorlinks   = true,
  urlcolor     = blue,
  linkcolor    = blue,
  citecolor   = blue 
}
\usepackage{xcolor}

\begin{document} 

   \title{2D and 3D unified RHD atmospheres and wind models of early B-supergiants including approximate NLTE effects}
   %\subtitle{Approximate NLTE effects on the ionization and wind dynamics}
  
\author{
D. Debnath\inst{1}\thanks{Corresponding author \email{dwaipayan.debnath@kuleuven.be}} 
\and J.O. Sundqvist\inst{1}
\and N. Moens\inst{2}
}

\institute{
Institute of Astronomy, KU Leuven, Celestijnenlaan 200D, 
3001 Leuven, Belgium
%\and
%Royal Observatory of Belgium, 1180 Brussels, Belgium
\and
Centre for Mathematical Plasma Astrophysics, Department of Mathematics, 
KU Leuven, Celestijnenlaan 200B, 3001 Leuven, Belgium
}

   \date{Received XXX; Accepted XXX}

   \titlerunning{Multi-D unified models of early B-supergiants}
   \authorrunning{Debnath et al.}

  \abstract
   % context  
   {Massive and luminous B supergiants are evolved massive stars with low surface gravities and lower effective temperatures, with substantial radiation-driven winds. Observationally, these stars display significant non-rotational broadening, stochastic low-frequency variability, and line-profile variability. Generally, these stars have been studied assuming 1D spherical symmetry.} 
   % aims
   {We present the first multi-D unified atmosphere and wind models for B supergiants, especially investigating the effects of approximate NLTE ionization and subsurface turbulence on the atmosphere and wind dynamics.}
  % methods heading (mandatory)
   {We computed multi-D box-in-a-star radiation-hydrodynamical unified atmosphere and wind simulations of three two-dimensional (2D) and one three-dimensional (3D) early B supergiant models using a hybrid opacity approach. We calculated line opacities from approximate NLTE populations, including line scattering. We calculated the flux-, energy-, and Planck-mean line opacities separately from our database consisting of $\sim 4$ million lines using Sobolev escape probabilities assuming two-level atoms.}
  % results heading (mandatory)
   {The models develop a highly structured and turbulent atmosphere and outflow, with time-averaged turbulent velocities of $\approx 40 - 55 \,\rm km/s$ at the photosphere in the 2D simulations, whereas the 3D model has $\varv_{\rm turb} \approx 25 \, \rm km/s$ at the photosphere. We found that luminosity variations in the multi-D simulations generally show excess power at lower temporal frequencies. In the cooler B1 and B1.5 models, the wind acceleration is more gradual than in the hotter B0 model. This leads to a significantly higher maximum wind speed in the B0 model than in the B1 and B1.5 models. Further, the approximate NLTE formalism strongly alters the ionization stratification in the wind compared to our earlier LTE formalism, modifying the line force. Mass-loss rates obtained in the models are of order $10^{-6} - 3\times10^{-6} \, \rm M_\odot/yr^{-1}$, and outer wind velocities are around 400 km/s for the B1 models and above 800 km/s for the B0 model. We further find that including turbulent pressure in hydrodynamically consistent NLTE co-moving frame steady-state 1D simulations of B supergiants with comparable parameters can produce higher mass-loss rates and lower terminal wind velocities than in corresponding co-moving frame models without $\varv_{\rm turb}$, in overall qualitative agreement with our multi-D simulations.}
  %  conclusions heading 
   {Our calculations suggest that in the lower surface gravity B supergiants, in addition to modifying the photospheric density scale heights, turbulent pressure also influences global properties of the wind. Due to the combination of strong ionization stratification and subsurface turbulence, a highly structured and variable photosphere and outflow is formed in these simulations, characterized by higher mass-loss rates and lower terminal wind speeds for the cooler B1, B1.5 and $\rm B1^{3D}$ models.}
   

   \maketitle
   
\section{Introduction}
\label{introduction}

B supergiants (BSGs) are evolved post-main-sequence massive stars characterized by high luminosities and relatively low surface gravities \citep{crowther_2006, langer_2012}. Their atmospheres are coupled to strong radiation-driven outflows, resulting in substantial mass loss \citep{pauldrach_1986, puls_2008}. Their effective temperatures, which are lower than those of O stars, place them in a regime where changes in ionization balance can effectively modify the line ensemble responsible for line driving \citep{Kudritzki_1999, petrenz_puls_2000, crowther_2006, lefever_2007, markova_2008, puls_2008, Searle_2008}. At their high luminosity-to-mass ratios, these stars are susceptible to dynamical motions associated with subsurface opacity zones  \citep{cantiello_2009, Grassitelli_2015, cassie_25, Ma_2026}. Furthermore, observations show large non-rotational line broadening and spectroscopic variability \citep{markova_2008} in both wind and photospheric lines \citep{simon_2017, haucke_2018, de_Burgos_2024, nadya_2024}, pointing to a highly dynamical atmosphere and outflow. Moreover, photometric analysis using space-based instruments such as \textsc{TESS} \citep{Ricker_2015} reveals a stochastic low-frequency variability (SLF) ubiquitously %\jon{really, ubiqitously? how many BSGs have been observed in SLF.. ?} \dwai{atleast in dominic's nature paper he says 111 (out of 114) of the K2 stars, and all 53 TESS LMC stars show significant SLF var} 
%Yes - my point was really specifically about BSGs - did he have them in sample?
%But you know this, so it's fine. 
present in BSGs \citep{Bowman2019, Ma_2024, Kourniotis_2025}. 

\begin{table*}
\caption{Fundamental parameters for the BSG models studied in this work}       
\label{tab:bsg_params}      
\centering          
\begin{tabular}{c | c c c c c c c c c c }  
\hline\hline       
Model & Dim & $M_\star/M_\odot$ & $R_\star/R_\odot$ &
$\log_{10} g_\star$ & $\varv_{\star}^{\rm turb} \rm [km/s]$ & $T_{\rm eff} \rm [kK]$ &
$\log_{10}(L_\star/L_\odot)$ & $\Gamma_{\rm e}$ & $\dot{M} \rm [M_\odot/yr]$ & $\varv_{\rm max} \rm [km/s]$ \\
\hline
B0     & 2D & 30 & 22.0 & 3.23 & 39 & 30.4 & 5.57 & 0.33 &$ 1 \times 10^{-6}$  & 802  \\
B1    & 2D & 30 & 32.2 & 2.89 & 55 & 25.6 & 5.60 & 0.35 & $3 \times 10^{-6}$  & 470\\
B1.5   & 2D & 25 & 36.9 & 2.70 & 50 & 22.7 & 5.51 & 0.34 & $3 \times 10^{-6}$  & 359\\
$\rm B1^{\rm 3D}$ & 3D & 30 & 29.5 & 2.97 & 25 & 26.5 & 5.59 & 0.34 & $3 \times 10^{-6}$ & 419 \\
\hline
\end{tabular}
\end{table*}

%, with terminal wind speeds adopted as input values from \citet{lamers_1995}.

\begin{figure*} 
\centering
\includegraphics[width=1.0\linewidth]{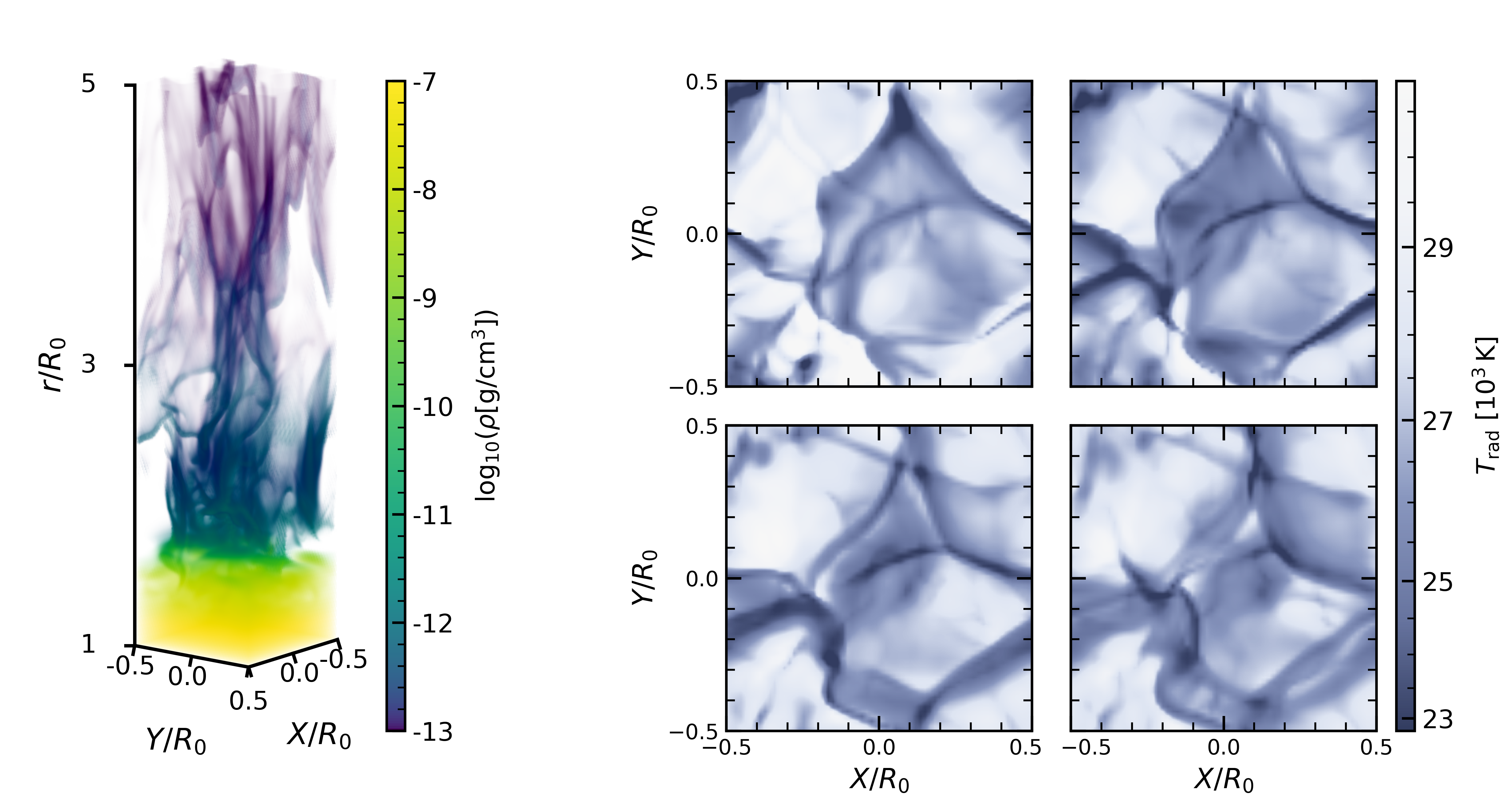}
\caption{The left figure illustrates the volume rendering of the density for the $\rm B1^{3D}$ model. The right panel shows the emergent monochromatic optical continuum intensity at $5000 \, \AA$ for four snapshots separated by $10 \, 000 \, \rm s$, viewed from the local normal surface ($\mu=1$). The colorbar shows the emergent radiation temperature calculated by $I_\nu = B_\nu(T_{\rm rad})$. An animation of this simulation is available at the following \href{https://youtu.be/ZTtym99vFFk?si=CT7SptmaHQzQbyHK}{online video}.}
\label{fig:bsg_volumerender} 
\end{figure*}

A long-standing uncertainty concerning line-driven winds in this regime is the so-called bi-stability jump. Early theoretical calculations suggested that luminous blue variables (LBVs) can have either a faster wind ($\sim 600 \, \rm km/s$) or a slow-moving wind ($\sim 200 \, \rm km/s$) \citep{pauldrach_puls_1990}. This transition was attributed to a change in the slope of the line-strength distribution, $\alpha$, around $20 \, \rm kK$ (see \citealt{cak1975}, \citealt{gayley_1995}, \citealt{puls_2000} for the definition of $\alpha$). This was later observed in terminal velocities around $21 \rm \, kK$ \citep{lamers_1995}. Monte Carlo and ionization calculations by \citet{Vink_1999} provided an explanation based on a change in ionization stage. As the effective temperature decreases, iron recombines from Fe-\textsc{iv} to Fe-\textsc{iii}, increasing the number of driving lines, with a subsequent increase in the mass-loss rate \citet{vink_2000, vink_2001}. Subsequent studies have suggested that this transition in terminal velocity may be more gradual, with large scatter \citep{crowther_2006,markova_2008,hawcroft_2023}. By contrast, the predicted increase in mass-loss rate at the bi-stability region has not been observed in recent empirical studies \citep{de_Burgos_2024, verhamme_2024, matheus_2025, verhamme_2026}.  
Theoretical global NLTE wind models using the \textsc{METUJE} code \citep{kriticka_2017} by \citet{kriticka_2021} do not predict an abrupt discontinuity in mass-loss rates and terminal wind speeds at the bi-stability, but rather a broader transition. The model grid by \citet{bjorklund_2023}, based on the dynamically consistent NLTE comoving-frame (CMF) method using \textsc{FASTWIND} by \citet{sundqvist_bjorklund_2019}, also does not generally predict an increase in mass-loss rate in that regime. However, recent hydrodynamically consistent model calculations by \citet{vink_2026} using \textsc{PoWR}  do recover a bi-stability jump. At higher Eddington ratios, the dynamically consistent OBA hypergiant \textsc{PoWR} models of \citet{Bernini-Peron_2026}, which include turbulent pressure, show complex mass-loss behavior across the temperature regime. These global wind predictions, such as mass loss and terminal velocities, are also quite sensitive to changes in the ionization balance, which can alter the line ensemble. In particular, $\alpha$ can change substantially depending on the dominant ionization stage \citep[][]{puls_2000, luka_2022, sundqvist_lime}. Such ionization stratification is well established in NLTE calculations of Wolf-Rayet (WR) winds \citep{lucy_1993, grafener_2005, sander_2020}. With their low effective temperatures, ionization stratification in the outflowing regions may become increasingly important in BSGs. 

%\jon{What is this about? has he done hydro-consistent models now, that I am not aware of?} \dwai{yeah its hydro consistent with powr, and they claim they even find bs at low gammae, however there is a caveat tho, they use vturb of 70km/s for all models, also some 21 km/s show some bs} 
%For higher-surface-gravity O stars, the modified density stratification does not strongly influence global mass-loss rates. \jon{I think we should take away this sentence about our O-star models actually, since I am no longer convinced about it..} 

%\jon{not sure I understand, I thought neither Sergio nor nadya had BSGs in their sample? 
%(But I may be wrong, but please check.)} \dwai{they both have bsgs in their sample, sergio's 2017 paper has stars above 15msun} 
%Ok, I thought it was only early objects, but that's fine then.

At the same time, previous multi-D radiation-hydrodynamical (RHD) simulations of O stars \citep{jiang_2015, schultz_2022, Debnath24, cassie_25, lara25, nico_2025} and quiescent LBVs \citep{pieter_2026} have shown significant velocity fields around the photosphere, except when very strong magnetic fields are present \citep{asif_2025}. Such turbulent velocity fields provide additional pressure support, modifying the density stratification around the photosphere \citep{Debnath24, gonzalez_2025, verhamme_thesis_2025}. Recently, \citet{Ma_2026} computed a 3D RHD envelope simulation of a BSG and found that instabilities associated with the subsurface iron-opacity bump generate a velocity field consistent with observed macroturbulent velocities \citep{simon_2017, nadya_2024}. This trend agrees with previous multi-D RHD O star modeling efforts shown in \citet{schultz_2020}, \citet{Debnath24}, \citet{lara25} and \citet{nico_2025}. Similar to O stars, atmospheres of BSGs are tightly coupled to the outflowing line-driven wind, and both regimes should be evolved in a unified way. Furthermore, we include line scattering and approximate NLTE (aNLTE henceforth) effects while calculating the opacities that control both momentum and thermal coupling between gas and radiation. 

We present the first multi-D unified atmosphere with line-driven wind models of early BSGs. In Sect.~\ref{sec:rhd_formulations} we summarize the RHD setup; Sect.~\ref{sec:model_setup} introduces and discusses the models. Sect.~\ref{sec:bsg_lineforce_parameters} describes how the line-force parameters alter between the LTE and aNLTE treatment. Sect.~\ref{sec:atm_wind_dyn} describes the velocity structure, mass-loss rates, and temporal variations. In Sect.~\ref{sec:1d_comp} we compare averaged multi-D wind properties with hydrodynamically consistent steady-state 1D CMF models. Finally, Sect.~\ref{sec:discussion} discusses observational implications, limitations, and concludes the work.

%For the relatively low-surface-gravity BSGs, significantly more material may be suspended in regions where mass loss is initiated. Thus, turbulence may not only modify the density scale height but also influence the global mass flux that is supplied to the line-driven outflow.  

%Additionally, it is unclear whether iron-group elements are solely responsible for initiating mass loss in these stars, especially in low-metallicity regions, as shown in \citet{sundqvist_lime} and \citep{verhamme_2026}. Recent calculations have also shown that the relative contributions of Fe and lighter elements near the critical point can vary considerably with effective temperature and metallicity \citep{kriticka_2025, verhamme_thesis_2025, verhamme_2026}. 

\section{RHD formulation and model construction}
\label{sec:rhd_formulations}

Following \citet{nico_2022a}, we solve the RHD equations on a finite volume mesh in multiple dimensions, using a Cartesian box-in-a-star setup including corrections for spherical divergence \citep{sundqvist_2018, nico_2022a} using the general-purpose hydrodynamic code \textsc{MPI-AMRVAC}\footnote{for more info check \href{https://amrvac.org/}{https://amrvac.org/}} \citep{xia_2018, keppens_2020}. We compute the mass, momentum, and both gas and radiation energy conservation equations, including effects of the radiation field in both momentum transfer and energy exchange. Gravitational force is included assuming a point mass. The system of equations is closed with the ideal gas law, assuming a constant adiabatic index $\gamma_g = 5/3$ and the flux-limited diffusion approximation \citep[FLD, ][]{levermore_pomraning_1981, nico_2022a}. Following \citet{Debnath_2026}, we separately calculate the flux-weighted mean that couples momentum transfer between radiation and matter and the energy- and Planck-weighted mean, which controls radiation-matter energy exchange. To evaluate the energy-exchange opacity means, we distinguish between line scattering and true absorption, taking aNLTE populations into account as outlined in \citet{Debnath_2026}. Line opacities are calculated using the \textsc{Munich Atomic Database} \citep{pauldrach_1998, puls_2000}, consisting of $\sim 4$ million lines for all the relevant elements (H-Zn) up to their eighth ionization stage. 

%\jon{Mention large data base here, as this remains an important point (- it's not like we do it for "10 typical liens" or something...). "Line opacities are calcualted fro mthe munich... consisting of .. X million number of .. for all relevant elements ... in suitable ionization ...}  

%However, we may still count the line-scattering part present in the OPAL tables \citep[see discussion in][]{Debnath_2026}. 
%unnecessary here.. (jon comment) 
%\jon{Ok, but again you are NOT mentioning the most important part - we're actualy computing this from our line data base including scattering and aNLTE effects now.. THIS is the key difference, not the fitting or other stuff. So I'd focus a bit more on that, and perhaps remove some other redundant stuff if needing space.}

For both the Planck-, energy-, and flux-mean opacities, we use the hybrid opacity technique introduced by \citet{luka_2021}, where we obtain the total opacities by summing the Rosseland means obtained from OPAL tables \citep{iglesias_1992, opal} with the calculated line opacities in a supersonic medium, assuming the \citet{sobolev_1960} approximation. Identical to \citet{Debnath_2026}, for the energy and Planck mean calculations we subtract electron scattering from the OPAL opacities, as coherent scattering does not contribute to energy exchange \citep{mihalas_1978}. We then fit the line-opacity multipliers using fit parameters first introduced by \citet{gayley_1995}. For the flux means, we follow \citet{luka_2022}, and for the energy and Planck means we follow \citet{Debnath_2026}. We use metal mass fraction $Z=0.02$ and helium mass fraction $Y=0.28$ for computing both line opacities and Rosseland means. We kept the initial and boundary conditions identical to \citet{Debnath24} and \citet{nico_2025}. For further details, we refer to \citet{nico_2022a}, \citet{nico_2022b}, \citet{Debnath24} and \citet{Debnath_2026}. 

We construct a small set of BSG models, namely B0, B1, B1.5, and $\rm B1^{3D}$, with the adopted stellar parameters summarized in Table~\ref{tab:bsg_params}. Each 2D model spans $8 \rm \, R_0$ radially, and laterally, the B0 model ranges over $0.4 \rm \, R_0$. Following \citet{Debnath24}, we define $R_0$ as the radius where the radiation temperature $T_{\rm rad} \simeq 500 \, \rm kK$. To ensure we adequately resolve atmospheric structures, we preemptively increased the lateral domain size compared to the previous O star simulations shown in \citet{Debnath24}, \citet{nico_2025} and \citet{lara25}, since the characteristic size of atmospheric structures generally correlates with the local scale height for O stars (Van der Sijpt et al. in prep). The cooler BSGs have low surface gravity and extended envelopes; this may lead to a large atmospheric pressure scale height and larger structures. To allow structure to develop laterally, we have extended the lateral range to 8 $R_0$ for the B1 and B1.5 models \footnote{Although this is broader than the physical circumference of the star $2 \pi R_0 \simeq 6.26 R_0$, this doesn't really have any effect on the physics within the simulation. Due to the pseudo-planar geometry used in these simulations \citep[See][]{nico_2022a}, the dynamics should be considered as local phenomena.}. For computational feasibility the 3D $\rm B1^{3D}$ model spans $4 \rm \, R_0$ radially, and $1 \rm \, R_0$ laterally. The 2D models resolve $0.004 \, R_0$ in both radial and lateral directions throughout. At minimum, one total pressure scale height is resolved by six grid cells, with similar resolution for the other 2D models. For the 3D model, we resolve down to $ 0.0156 \, R_0$, for which at least one grid cell length resolves one total pressure scale height. 

\section{Characteristics of the early BSG models}
\label{sec:model_setup}

\begin{figure} 
\centering
\includegraphics[width=1.0\linewidth]{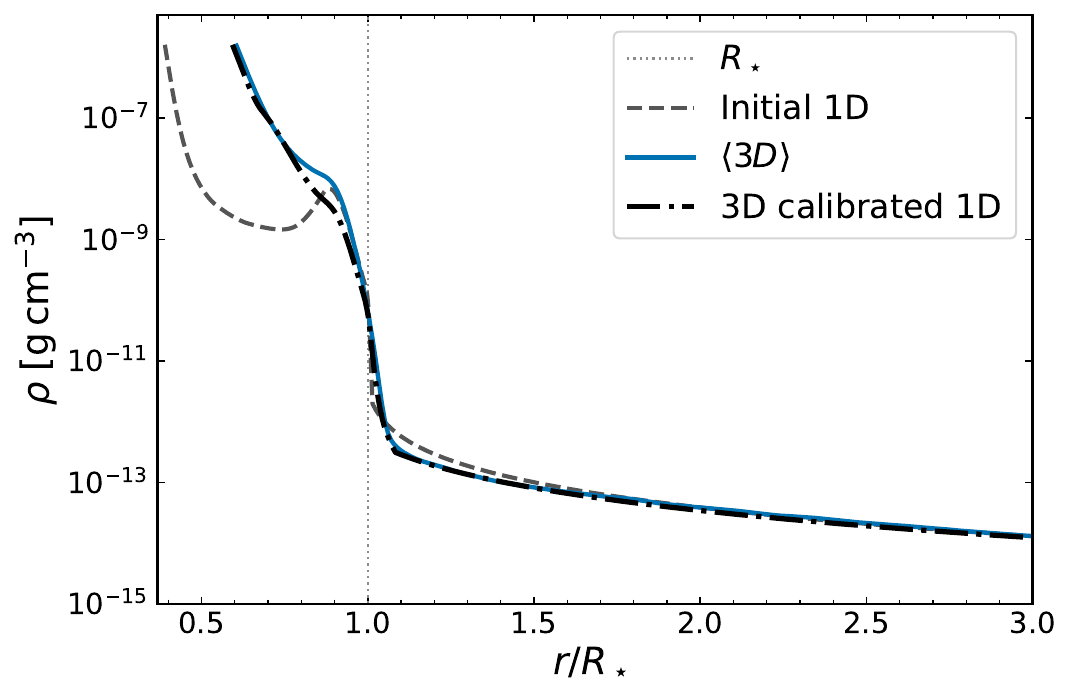}
\caption{The figure displays the temporally and laterally averaged 3D density as a function of $r/R_\star$, along with the initial 1D structure and the modified density structure obtained using a turbulent velocity $\varv_{\rm turb} = 24 \, \rm km/s$ inferred from the 3D simulation and $\alpha_{\rm MLT} = 2.0$. The vertical dashed line indicates the effective temperature of the star at the photosphere.}
\label{fig:bsg_density_t} 
\end{figure}

% The B1 model has radially and temporally averaged turbulent velocities of $55 \, \rm km/s$ at the photosphere, whereas the B0 and B1.5 models have averaged turbulent velocities of 39 $\rm km/s$ and 50 $\rm km/s$, respectively. Similar to the previous 3D O star trends in \citet{lara25}, the $\rm B1^{3D}$ model has a much lower photospheric turbulent velocity, around $24 \, \rm km/s$. 
%Radiation pressure dominates throughout the simulation domain, with gas pressure becoming as important as the radiation pressure only in the innermost layers. 

% \jon{Missing fig: some nice volume rendering or something of a 3D model snapshot. should this not be included, I think you should "show-case" this at least? (Also, at some position we should put a footnote with link to our surface brightness movie, for the earxiv-version. We can take the youtube one for course, or perhaps prepare one for home-page as well, and link to that so people can download.}

After an initial relaxation phase, the models settle to a quasi-stationary regime similar to \citet{nico_2022b}, \citet{Debnath24} and \citet{nico_2025}. As in O stars, structure appears in deep subsurface layers around the iron-opacity bump \citep{jiang_2015, Debnath24, cassie_25} and then propagates outward to the photosphere and outflowing regions, as shown in Fig.~\ref{fig:bsg_volumerender} (left panel) for the $\rm B1^{3D}$ model. This creates highly turbulent layers around the photosphere. 
On average, the 2D models have higher laterally and temporally averaged photospheric turbulent velocities than the 3D model. Such high turbulent velocities result in turbulent pressure of the same order as radiation pressure in layers above the iron-opacity bump and the photosphere. Qualitatively, the radiation, gas, and turbulent pressure behave similarly to the O stars shown in \citet{Debnath24} and \citet{nico_2025}. 

Following \citet{lara25}, Fig.~\ref{fig:bsg_volumerender} (right panel) plots the emergent monochromatic continuum intensity at $5000 \, \AA$ for a few representative snapshots for the $\rm B1^{3D}$ model. The emergent specific intensity, $I_\nu$, is computed by solving along rays perpendicular to the stellar surface ($\mu=1$), assuming a source function given by the Planck function $B_\nu(T_{\rm gas})$, where $T_{\rm gas}$ is the local gas temperature. We use the local Rosseland mean to compute the opacities, $\kappa_\nu = \kappa_{\rm R}$. The emergent intensities illustrate layers around the photospheric optical depths of order unity. We then express the radiation temperature, $T_{\rm rad}$ as $I_\nu \equiv B_\nu(T_{\rm rad})$. Fig.~\ref{fig:bsg_volumerender} displays spatial variations of the emergent intensities with large-scale bright and dark patches, with the local radiation temperature varying by a few thousand Kelvin around the mean. However, compared to the O4 model in \citet{lara25}, the structures here are much larger, with typical scales visible in the figure of order $\sim 8 \, R_\odot$. 

Due to their higher turbulent pressure, the 2D models are more inflated than the 3D model. The photospheric radius is $\sim 8\%$ smaller in the $\rm B1^{3D}$ model compared to the 2D B1 model. As a result, the 3D model has a higher effective temperature than the 2D model. However, the 3D simulation has a smaller spatial resolution than the 2D model. Previous resolution tests in 2D have shown that atmospheric inflation slightly decreases as the resolution is increased. This suggests that a higher 3D resolution would likely exhibit a slightly smaller photospheric radius.
 
At maximum, the radiative enthalpy (radiation advection) \citep[see][for definition]{Debnath24} carries only $10\%$, $20\%$, and $23\%$ of the total energy for the 2D B0, B1, and B1.5 models, respectively. Like its 2D counterpart, the $\rm B1^{3D}$ model also carries a maximum of $20\%$ of the total energy by radiative enthalpy, which is of a similar order of magnitude as the O supergiant models in \citet{nico_2025}. This is much lower than in transitional cool-plate models with higher luminosity-to-mass ratios in \citet{pieter_2026}, which find that radiation enthalpy carries $\sim 90\%$ of the total energy at maximum. For both 2D and 3D B1 models, the region where radiation enthalpy contributes at least $1\%$ of the total flux extends from $230 \, \rm kK$  to  $80 \, \rm kK$. This is somewhat broader than for the O4 model in \citet{Debnath24} and \citet{lara25}, which extends $ 220 - 130 \, \rm kK$.

\begin{figure} 
\centering
\includegraphics[width=1.0\linewidth]{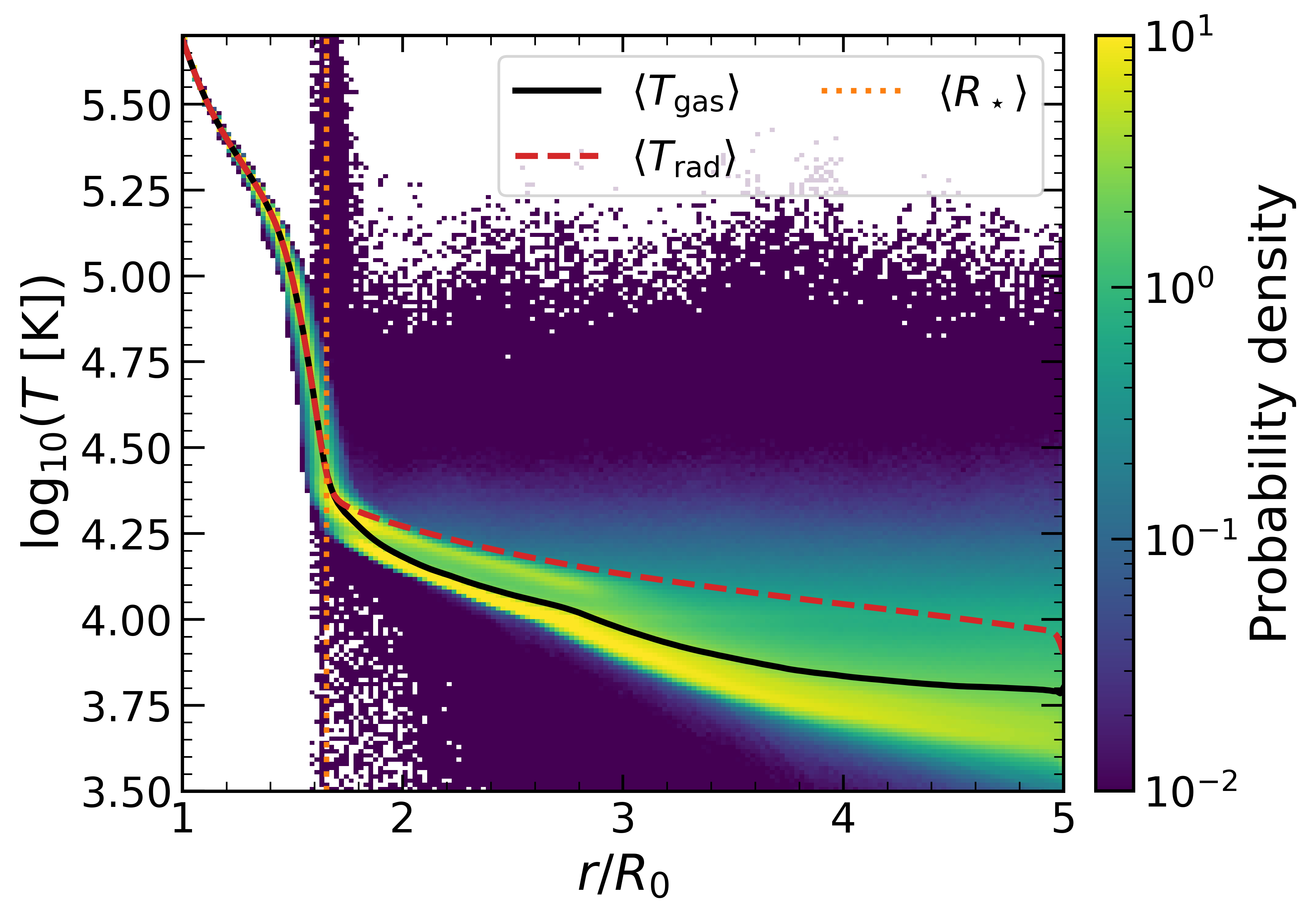}
\caption{Probability density distribution of the gas temperature, with yellow indicating high probabilities and blue vice versa for the $\rm B1^{3D}$ model using one hundred 3D snapshots. The black solid curve and red dashed curve indicate the temporally and laterally averaged gas and radiation temperature, respectively. The orange vertical line shows the average photosphere.}
\label{fig:bsg_tgas} 
\end{figure}

Figure~\ref{fig:bsg_density_t} displays the laterally and temporally averaged density profile for the $\rm B1^{3D}$ model, along with the initial 1D conditions and a 1D model calculated with 3D-informed turbulent velocities and radiative enthalpy transport. To obtain a shallower density stratification in the inner layers around the iron-opacity bump, following \citet{Debnath24}, we included convection using a simple \citet{schwarzschild_1958} criterion with mixing-length $\alpha_{\rm MLT} = 2$, since it provides a better fit to 3D data compared to $\alpha_{\rm MLT} = 1.5$ previously used in \citep{nico_2025} for O stars, together with a turbulent pressure support characterized by a turbulent velocity $\varv_{\rm turb} = 25 \, \rm km/s$. We kept $\varv_{\rm turb}$ constant throughout the photosphere and subphotospheric layers until $T_{\rm rad} = 60 \rm \, kK$. At higher temperatures, we gradually reduce the turbulent velocity until $T_{\rm rad} = 200 \rm \, kK$, beyond which a substantially stronger inward taper is applied. As in the 3D model, the calibrated 1D model shows turbulent pressure exceeding gas pressure in the subphotospheric layers. Although radiation pressure remains the dominating contributor in magnitude, turbulent pressure provides a substantial additional contribution to the atmospheric pressure stratification. Further, in layers around the iron-opacity bump, the 1D model shows convection carrying $\sim 58 \%$ of the total energy flux at maximum. 

%\jon{Add how much vturb contribute to the hydrostatic balance as well in the 1D model here, higher or lower than Prad, Pgas?}

In the dense atmosphere, $\kappa_E$ and $\kappa_B$ are approximately equal to each other. Further out, $\kappa_E$ and $\kappa_B$ begin to differ, and the gas and the radiation temperatures begin to decouple. On average, in these regions, the gas stays cooler than the radiation temperature, as shown in Fig.~\ref{fig:bsg_tgas} for the $\rm B1^{3D}$ model. However, small pockets of hot gas are present where the material is compressed at the shock fronts, heating the gas to temperatures higher than the local radiation temperature. The volume of gas in the outflowing regions that stays hotter than the local radiation temperature is $3 \%$, and $4.9 \%$ for the B1 and $\rm B1^{3D}$ models, respectively. However, for both B1 and $\rm B1^{3D}$ models, the volume of gas that stays above the effective temperature of the star is $0.2\%$, which is lower than that of the O stars shown in \citet{Debnath_2026}. 

%Qualitatively, this behavior follows the O-dwarf model in \citet{Debnath_2026}. 

%Changes in ionization modify the line force and therefore influence the outflow. We therefore first examine the line-force parameters across the computational domain.  

\section{Ionization and line-force parameters}
\label{sec:bsg_lineforce_parameters}

The line-driven radiative acceleration depends on the ionization state of the gas \citep{cak1975, gayley_1995, puls_2000, sander_2020, luka_2022, sundqvist_lime}. The aNLTE formalism modifies the ionization stratification, which alters the wind dynamics. In this section, we discuss the effects of changes in ionization on the line ensemble.   

%\begin{figure*} 
%\centering
%\includegraphics[width=\linewidth]%{Pictures/line_params.png}
%\caption{Color maps of the line-force parameters and the corresponding effective line opacity for
%the B1 model. The left columns indicate the LTE and the right indicate the aNLTE calculations.}
%\label{Fig:bsg_lineparams} 
%\end{figure*}

\begin{figure} 
\centering
\includegraphics[width=1.0\linewidth]{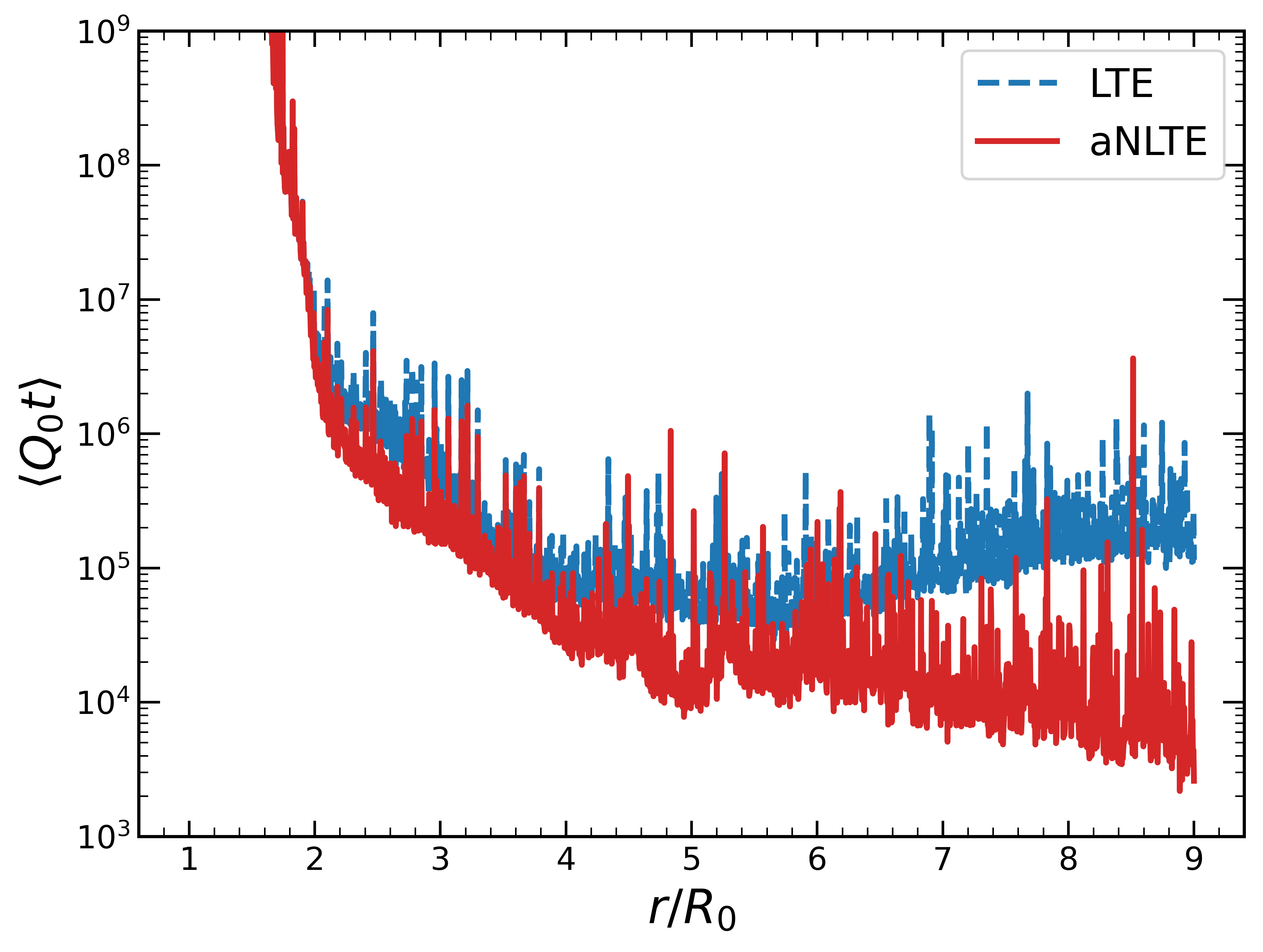}
\caption{Laterally and temporally averaged $Q_0t$ as a function of radius for the B1 model. The LTE and aNLTE calculations are shown by the blue dashed and red solid curves, respectively.}
\label{fig:bsg_Q0t} 
\end{figure}

\begin{figure*}
\centering
\includegraphics[width=\linewidth]{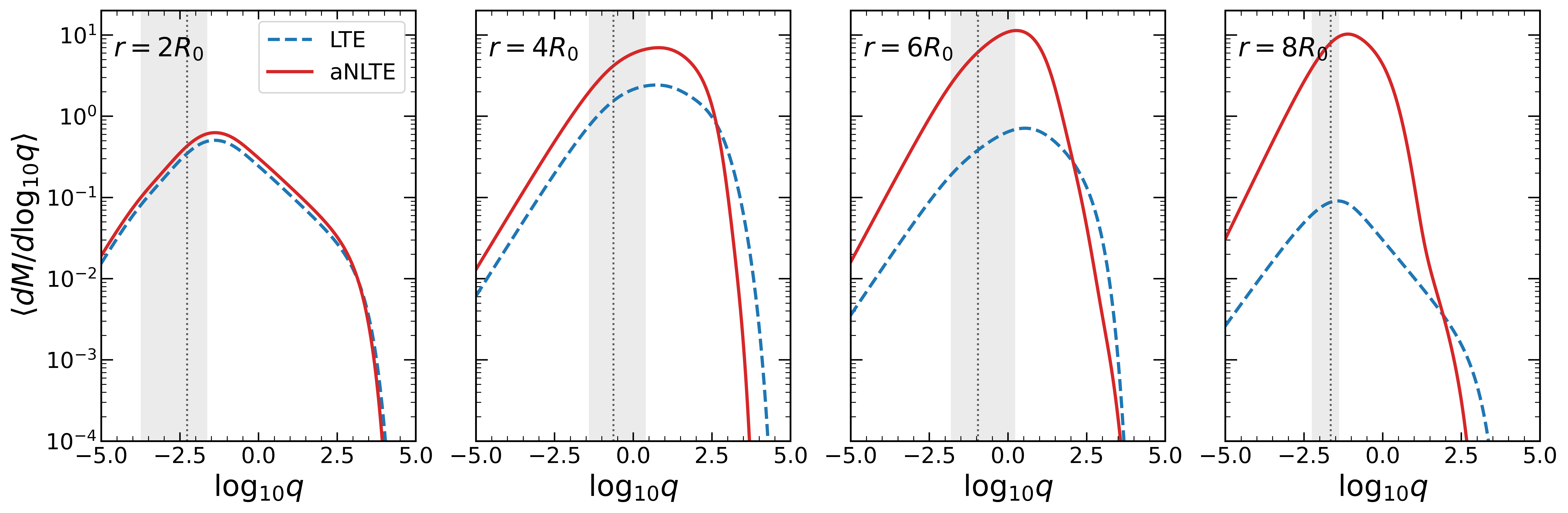}
\caption{The contribution of the logarithmic line-strength interval to the force multiplier at various radii for the B1 model for a laterally averaged singular snapshot. The LTE and the aNLTE distributions are shown by the curves indicated in the legend. The vertical dotted line shows the median value of $q=1/t$, for which the Sobolev optical depth is unity, and the shaded regions show the lateral variations for a snapshot.}
\label{fig:b1_dmdlogq}
\end{figure*}

\subsection{Line-strength distribution}

The aNLTE treatment modifies the level populations and thus the line-force parameters entering the flux-mean opacity and radiative acceleration \citep[see][]{Debnath_2026}. To isolate the aNLTE effects, we calculate reference LTE populations using the local gas density, $\rho$, while imposing $T_{\rm gas}=T_{\rm rad}$. Using $T_{\rm gas}$ from aNLTE would simultaneously introduce two effects, namely, modifications to the level populations and the thermal structure. In aNLTE, the level populations depend on both $T_{\rm gas}$ and $T_{\rm rad}$, $\rho$, and the dilution factor. For details, we refer to \citet{Debnath_2026} and references therein. In such a parameterization, the line-force multiplier is fitted using three parameters. $\bar{Q}$ provides its maximum value in the optically thin limit. $Q_0$ sets the upper cutoff of the line-strength distribution. $\alpha$ determines the slope of the power law \citep{gayley_1995, luka_2022}.

The 2D and the 3D models produce qualitatively similar line-force parameter behavior over the overlapping radial domain. However, due to its larger radial extent, we focus on the 2D model in the following analysis. We find that near the photosphere of the B1 model ($\sim 2 R_0$), the line-force parameters behave qualitatively similarly between the two methods; farther out, they differ. In dense wind pockets, $\bar{Q}$ becomes substantially larger in the aNLTE treatment; simultaneously, $Q_0$ decreases and $\alpha$ increases. Thus, aNLTE changes both the effective strength of the line ensemble and the shape of the line-strength distribution through $Q_0$ and $\alpha$. The dynamical importance of the line-force parameters also changes depending on the local characteristic depth $t = \kappa_{\rm e} \rho c / |{\rm d} \varv_{\rm r}/{\rm d}r|$ \citep{luka_2022, sundqvist_lime}. In the force-multiplier formulation, $Q_0$ enters through the combination $Q_0 t$, informing whether the line ensemble is approaching the optically thin or saturation limit. Guided by this, we examine the laterally averaged $Q_0t$ for the 2D B1 model in Fig.~\ref{fig:bsg_Q0t}. At the photosphere, $Q_0 t$ remains large in both methods and naturally decreases at larger radii, although values obtained for the present method are smaller than LTE. Nevertheless, over the entire computational domain we find that $Q_0t\gg1$, indicating that the wind on average does not approach the effective optically thin limit, $Q_0 t \ll 1$ \citep{sobolev_1960, cak1975, gayley_1995, luka_2022}. However, $Q_0 t \gg 1$ applies only to the average structure; in local pockets of rarefied gas, $Q_0 t$ may become small. Further, we point out that not every contributing line is optically thick. A line of strength $q$ with Sobolev depth $\tau_{{\rm S}}=qt$ would be optically thin when $qt\lesssim1$ and saturated when $qt\gtrsim1$. On average over the entire computational domain, the 
ensemble remains in the optically-thick regime.

\begin{figure*}
\centering
\includegraphics[width=\linewidth]{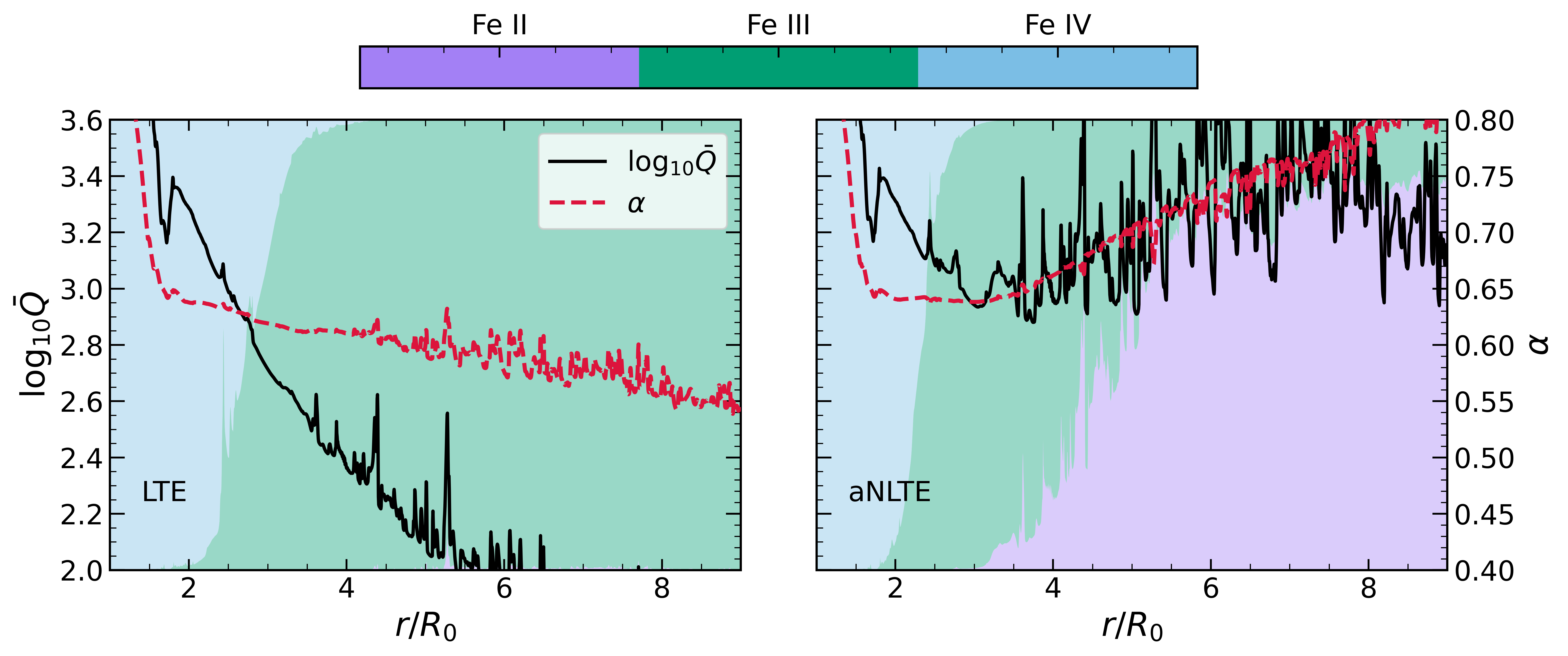}
\caption{Colors showing laterally averaged Fe ion fractions for the B1 model for a single snapshot. The ionization stages are indicated in the labels. The left and right columns show the LTE and aNLTE calculations, respectively. The red dashed curves show the corresponding laterally averaged $\alpha$, and $\log_{10}\bar{Q}$ is shown in black solid curves.}
\label{fig:b1_ion_qbar}
\end{figure*}

To understand how changes in $Q_0$ and $\alpha$ affect the line force, we examine how different line strengths contribute to the total force multiplier. Following \citet{gayley_1995,luka_2022}, the statistical distribution of the lines can be written as
\begin{equation}
{\rm d}N_\nu = f(\nu) \dfrac{\bar{Q}}{\Gamma(\alpha)Q_0^2}\left(\frac{q}{Q_0}\right)^{\alpha-2}
\exp\left(-\frac{q}{Q_0}\right)\,{\rm d}\nu\,{\rm d}q ,
\label{eq:bsg_line_distribution}
\end{equation}
Here, $f(\nu)=1/\nu$ provides a standard statistical frequency distribution of lines \citep{friend_castor_83,puls_2000,luka_2022}, and $\Gamma(\alpha) = \int_0^\infty x^{\alpha-1}\exp(-x)\,{\rm d}x$ is the Gamma function \citep{gayley_1995, luka_2022}. Also, Eq.~\eqref{eq:bsg_line_distribution} describes the number of lines within a given line-strength and frequency interval. The contribution from all lines to the force multiplier within some ${\rm d}\nu\,{\rm d}q$ is,
\begin{equation}
{\rm d}M = M_i(q,t)\,{\rm d}N_\nu .
\label{eq:bsg_line_multiplier}
\end{equation}
where $M_i(q,t)$ is the contribution from a single line. Substituting the value of the line distribution from Eq.~\eqref{eq:bsg_line_distribution} and the force multiplier shown in \citet{luka_2022} and then integrating Eq.~\eqref{eq:bsg_line_multiplier} over frequency, we arrive at the contribution to the force multiplier per logarithmic interval in line strength,
% \jon{Not sure I follow here actually? M is normally the summed force multiplier, or is not that here?} 
%
\begin{equation}
\frac{{\rm d}M}{{\rm d}\log_{10}q}  
= \ln(10) \frac{\bar{Q}}
{\Gamma(\alpha)Q_0t} \left(\frac{q}{Q_0}\right)^{\alpha-1}
\exp\left(-\frac{q}{Q_0}\right) \left(1-\exp[-qt]\right).
\label{eq:bsg_dmdlogq}
\end{equation}
Eq.~\ref{eq:bsg_dmdlogq} describes how lines within a logarithmic line strength interval contribute to the total force multiplier. When $qt\ll1$ and $q\ll Q_0$,
\begin{equation}
\frac{{\rm d}M}{{\rm d}\log_{10}q}
\propto q^\alpha.
\label{eq:m_logq_thin}
\end{equation}
As $\alpha > 0$, the contribution from strong lines to the force multiplier increases. In the opposite optically thick limit, as $qt\gg1$ and $q\ll Q_0$,
\begin{equation}
\frac{{\rm d}M}{{\rm d}\log_{10}q}\propto q^{\alpha-1}.
\label{eq:m_logq_thick}
\end{equation}
and with $\alpha<1$, the contribution to the force multiplier decreases with increasing line strength. The force contribution therefore peaks between the optically thin and thick transition, with $qt$ typically in orders of unity. The exact peak value, however, depends on $\alpha$ and the exponential cutoff $q/Q_0$. Once a line is highly saturated, its contribution only weakly depends on line strength; therefore, extending the line distribution towards larger $q$ will not significantly increase the force multiplier.

Fig.~\ref{fig:b1_dmdlogq} shows the distribution of ${\rm d}M/{\rm d}\log_{10}q$ across different line strengths for various radii for a laterally averaged B1 snapshot. The dotted line shows the line strength at which $q = 1/t$, corresponding to Sobolev optical depth $\tau_{\rm S} = 1$. Lines to the left of the dotted line have $qt<1$ and are optically thin, whereas lines on the right have $qt > 1$ and are optically thick. Near the photosphere, the left panel of Fig.~\ref {fig:b1_dmdlogq} (at $r = 2 R_0$) shows that LTE and aNLTE distributions are almost identical in shape and magnitude, agreeing with our previously found identical line-force parameters in these regions. At $r = 6R_0$ and $8 R_0$, these low- and intermediate-strength lines contribute substantially more to the total force multiplier in aNLTE than in LTE, which in turn produces a larger line-force. At the same time, aNLTE at $r = 8 R_0$ extends to a smaller high-q value than LTE. However, these high-q lines contribute almost negligibly to the total force multiplier. Therefore, the present aNLTE treatment in the outer wind shifts the effective line-strength distribution to a larger contribution from lower- and intermediate-strength lines, resulting in an increased total integrated line force.

\subsection{Ionization stratification}
\label{ion_stratification}

The differences in the line-strength distribution can be understood by examining the ionization structure in the wind. Fig.~\ref{fig:b1_ion_qbar} compares the laterally averaged iron ion fractions for a single snapshot with the overplotted line-force parameters $\alpha$ and $\bar{Q}$ for LTE and aNLTE formalisms. A qualitatively similar ionization stratification is found in the $\rm B1^{3D}$ model; however, as discussed earlier, its more limited radial extent does not allow us to capture the full transition in the outer wind, and thus we focus on the B1 model.

As can be seen from Fig.~\ref{fig:b1_ion_qbar} in LTE, Fe-\textsc{iii} remains the dominant ionization stage throughout the wind. Similarly, in the present method, Fe-\textsc{iii} dominates around the photosphere ($\sim 2 R_0$) and the wind-launching region; however, further out, it recombines to Fe-\textsc{ii}. Consequently, in LTE, $\bar{Q}$ is substantially lower in the outer regions, whereas in the present aNLTE method $\bar{Q}$ stays large throughout the outflowing regions of this simulation domain. The regions where $\bar{Q}$ and $\alpha$ diverge between the two methods coincide with the region where we find substantial recombination to Fe-\textsc{ii} in the present method, indicating that recombination to Fe-\textsc{ii} could alter the line ensemble, resulting in different $\bar{Q}$ and $\alpha$ values for the present method. Although CNO ions also undergo significant recombination in the aNLTE, it occurs further inwards compared to where the line-force parameters diverge.

%We now examine the dynamical properties of the BSG outflows, focusing on the velocity structure, mass-loss rates, and their temporal variations. 
%\section{Discussion}
%\label{sec:dicussion}

\section{Atmospheric and wind dynamics}
\label{sec:atm_wind_dyn}

We now examine the dynamical properties of the BSG outflows, focusing on the velocity structure, mass-loss rates, and their temporal variations.

\subsection{Velocity structure}
\label{sec:bsg_vel_str}

\begin{figure}
\centering
\includegraphics[width=1\linewidth]{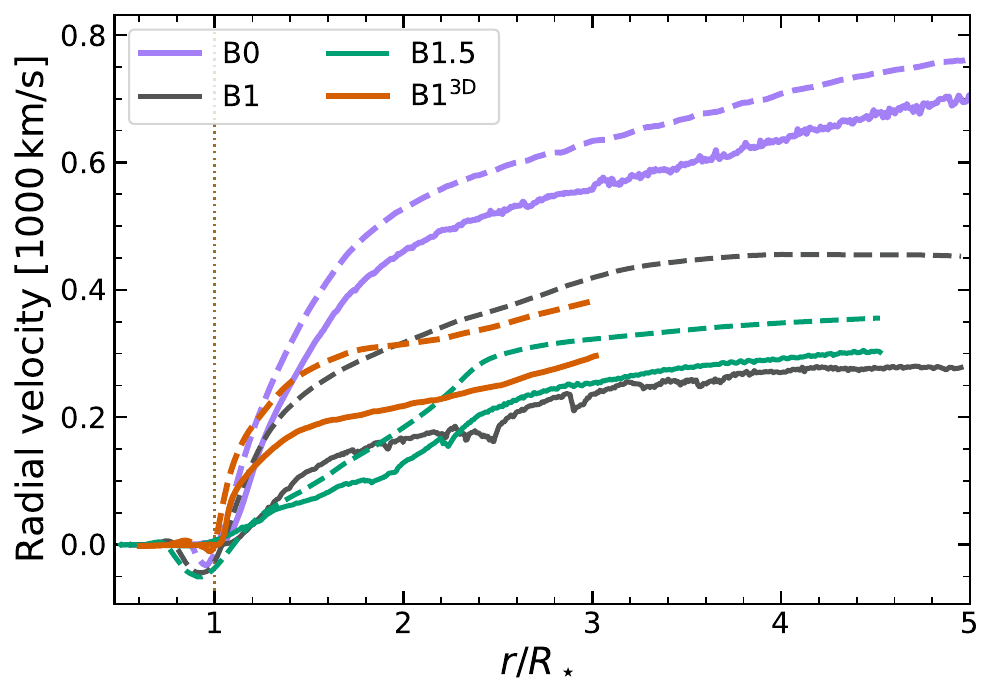}
\caption{Dashed curves present laterally and temporally averaged radial velocities, and solid curves denote density-weighted radial velocities, as a function of r/$R_\star$ over 60 snapshots for various models. The vertical lines indicate the average photosphere at optical depth $\tau = 2/3$.}
\label{fig:bsg_velocity_profiles}
\end{figure}

In Fig.~\ref{fig:bsg_velocity_profiles}, we show the laterally and temporally averaged and density-weighted radial velocities for the BSG models. The hotter B0 model has a rapidly accelerating wind. The mean radial velocity rapidly reaches a few hundred $\rm km/s$ outside the photosphere. The wind is still accelerating at the edge of the computational domain; however, the acceleration is much more gradual compared to the initial steep rise in velocity. On average, the wind reaches velocities up to $700\, \rm km/s$. In contrast, both cooler B1 and B1.5 models develop a qualitatively different velocity structure. The wind accelerates more gradually than in the B0 star and reaches much lower maximum wind speeds, only a few hundred $\rm km/s$ at the edge of the simulation domain. This may indicate a transition from a fast B0 wind to a substantially slow-moving outflow in the cooler B1 and B1.5 stars. The $\rm B1^{3D}$ model has a faster acceleration at the base of the wind; at maximum it reaches $\sim 300$ km/s, similar to the B1 and B1.5 stars. This slowly accelerating wind may be related to the classical change in wind velocity across wind recombination zones \citep{pauldrach_puls_1990, Vink_1999}, or perhaps to a different solution branch that under certain circumstances can be obtained within modified CAK wind theory \citep{Cure04,cure_11,venero_2016,venero_2024}. In any case, the emergence of our slower winds suggests an origin related to ionization changes in the wind. 
Interestingly, unlike its 2D counterpart, the $\rm B1^{3D}$ model does not exhibit a pronounced negative mean radial velocity around the photosphere; rather, the mean remains closer to zero, indicating that the strong photospheric infall found in the 2D models is significantly reduced in 3D. Additionally, Fig.~\ref{fig:bsg_velocity_profiles} also shows the systematic difference between the straight averages and their density-weighted counterparts. The density-weighted velocities remain below the ordinary mean velocities throughout the outflowing regions. The relatively low density-weighted velocities indicate that a large fraction of wind mass is slowly accelerating.

%However, unlike modified CAK studies, we do not prescribe a constant ionization parameter $\delta$ for the complete global wind; instead, the ionization balance determines local line-force parameters that govern the wind dynamics. Nevertheless, 
%possible physical connection to the ionization-driven transition 
%underlying the $\delta$-slow branch. 
%JS-NOTE: I tried to modify this, since it was not really coherent - you would need to really study these models to get to what they may, or may not, mean... (it is MUCH more subtle than this. something for another time I'd say.)

\begin{figure}
\centering
\includegraphics[width=1
\linewidth]{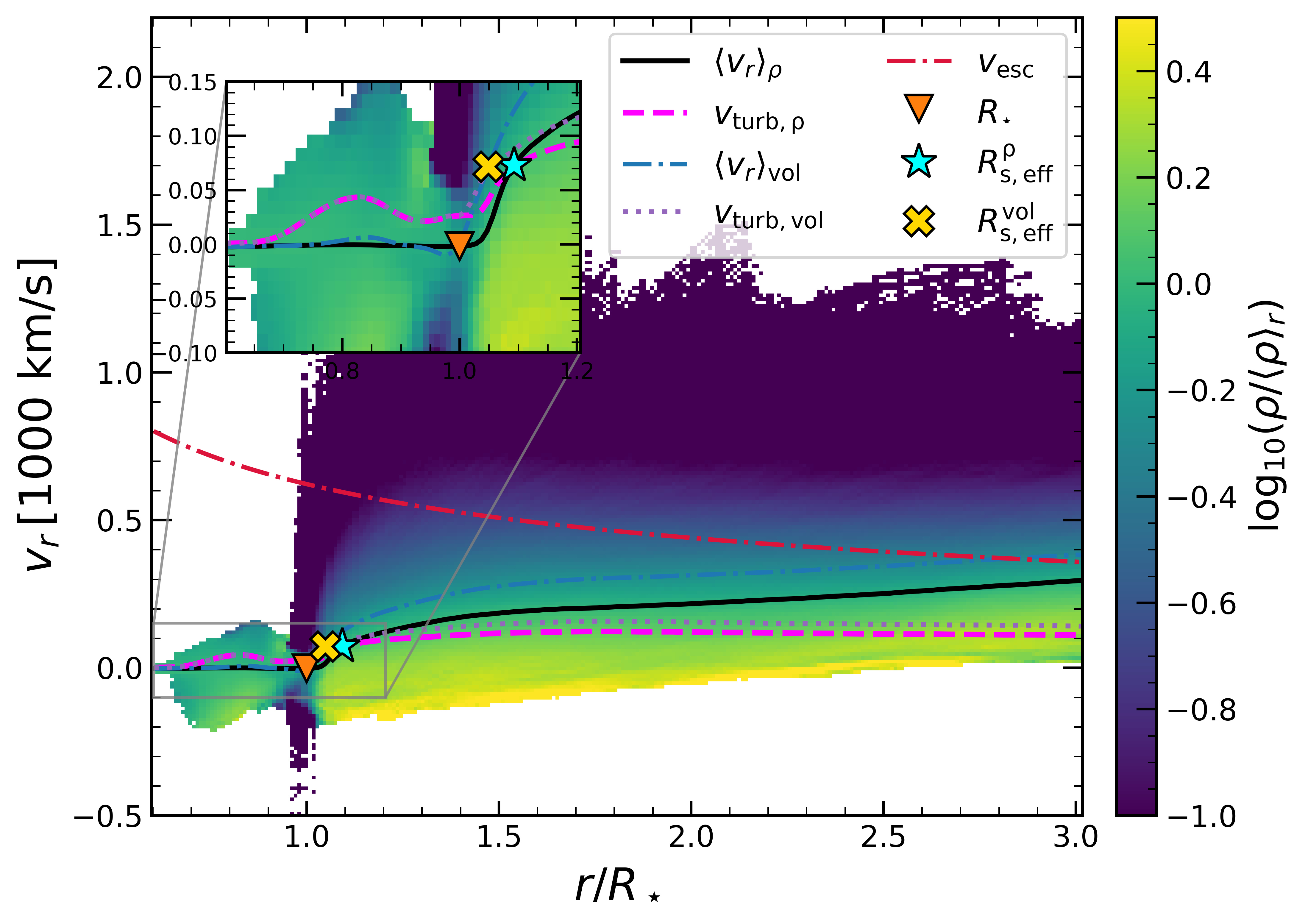}
\caption{Radial velocity distribution of the $\rm B1^{3D}$ model as a function of $r/R_\star$, using 100 snapshots. Colors represent the density contrast, with yellow representing the high-density medium and blue the rarefied parts. The red dashed-dotted curve is the local escape velocity. The orange triangle is the laterally and temporally averaged photosphere $R_\star$. The black solid line is the laterally and temporally averaged density-weighted radial velocity, and the blue dash-dotted line is the laterally and temporally averaged radial velocity. The magenta dashed line is the turbulent velocity calculated by subtracting the density-weighted mean radial velocity, whereas the purple dotted line is constructed by subtracting the simple mean from the radial velocity. The cyan star provides the effective sonic point, $R^{\rm \rho}_{\rm s, eff}$, calculated using $\varv_{\rm turb , \rho}$, and the golden cross provides the effective sonic point, $R^{\rm vol}_{\rm s, eff}$, obtained using $\varv_{\rm turb , vol}$. The inset provides a zoomed view of the inner regions.}
\label{fig:bsg_b1_velocity_distribution}
\end{figure}

%However, in the present approximate NLTE models of the BSGs, the dependence is more complicated due to the modified ionization balance and hence the local values of the force multipliers. 

\begin{figure*}
\centering
\includegraphics[width=1.\linewidth]{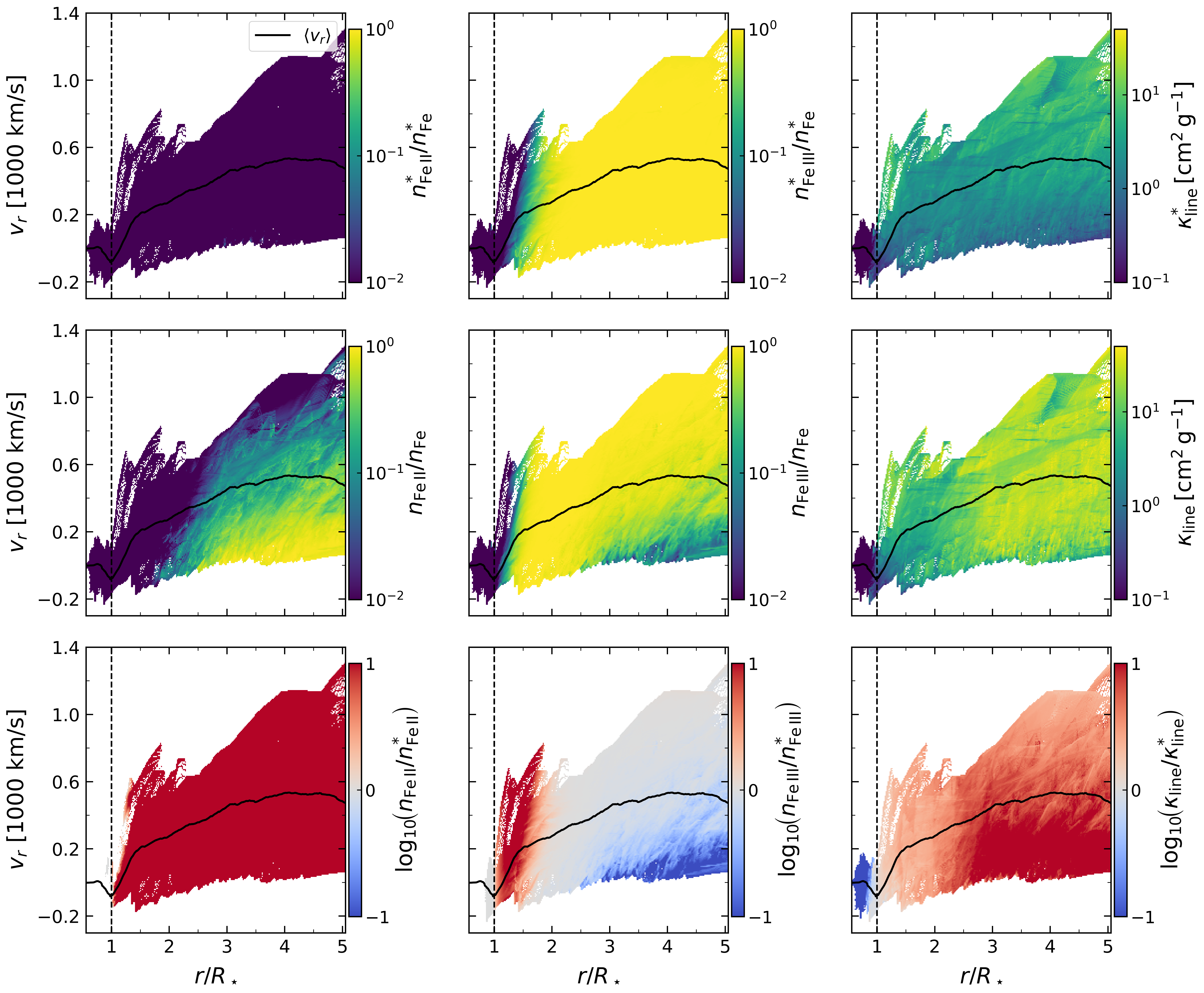}
\caption{The plots show the distribution of the radial velocities as a function of radius for a single snapshot for the B1 model. The asterisk here denotes the LTE calculations. The colors represent different quantities. In the top panel, from left to right, the LTE $n^*_{\rm Fe~\textsc{ii}}/n^*_{\rm Fe}$, $n^*_{\rm Fe~\textsc{iii}}/n^*_{\rm Fe}$, and the total line opacity in LTE, $\kappa^*_{\rm line}$. The middle row from the top, from left to right, provides the aNLTE $n_{\rm Fe~\textsc{ii}}/n_{\rm Fe}$, $n_{\rm Fe~\textsc{iii}}/n_{\rm Fe}$, and the total line opacity in aNLTE, $\kappa_{\rm line}$. The bottom row, from left to right, compares $n_{\rm Fe~\textsc{ii}}/n^*_{\rm Fe~\textsc{ii}}$, $n_{\rm Fe~\textsc{iii}}/n^*_{\rm Fe~\textsc{iii}}$ and $\kappa_{\rm line}/\kappa^*_{\rm line}$. The black curve provides the laterally averaged radial velocity.}
\label{fig:bsg_snapshot200_probdens}
\end{figure*}

Figure~\ref{fig:bsg_b1_velocity_distribution} shows the radial velocity distribution, with colors representing the density contrast for the $\rm B1^{3D}$ model. Although the mean flow, shown with black curves, in the outer line-driven regions moves at $\sim 300 \, \rm km/s$, the rarefied medium moves much faster, reaching velocities exceeding $1000 \, \rm km/s$. This anti-correlation of line-driving with density has also been observed in previous multi-D simulations of both WR stars \citep{nico_2022b} and O stars \citep{Debnath24}. The rarefied medium reaches escape velocity at a much smaller radius than the mean flow, which only reaches escape velocity beyond $\sim 2.5 \, R_\star$. In contrast, the 2D O4 model presented in \citet{Debnath24} shows that the mean flow already reaches escape speed at $1.68 \, R_\star$. In the present B1 models, the dense wind component barely drags along over an extended radial range, while the fast-moving material may have already attained escape velocity. Even at the outer edge of the simulation domain at $3.0 \, R_\star$, the dense medium stays much below the local escape speed. For the larger radially extended 2D B1 model, even at $4.5 \, R_\star$ the dense material moves at velocities lower than the local escape velocity. By contrast, in the O4 model in \citet{Debnath24}, at the outer edge of their simulation domain at $2.4 \, R_\star$, even the dense medium attains the local escape speed. 

\begin{figure*}
\centering
\includegraphics[width=\linewidth]{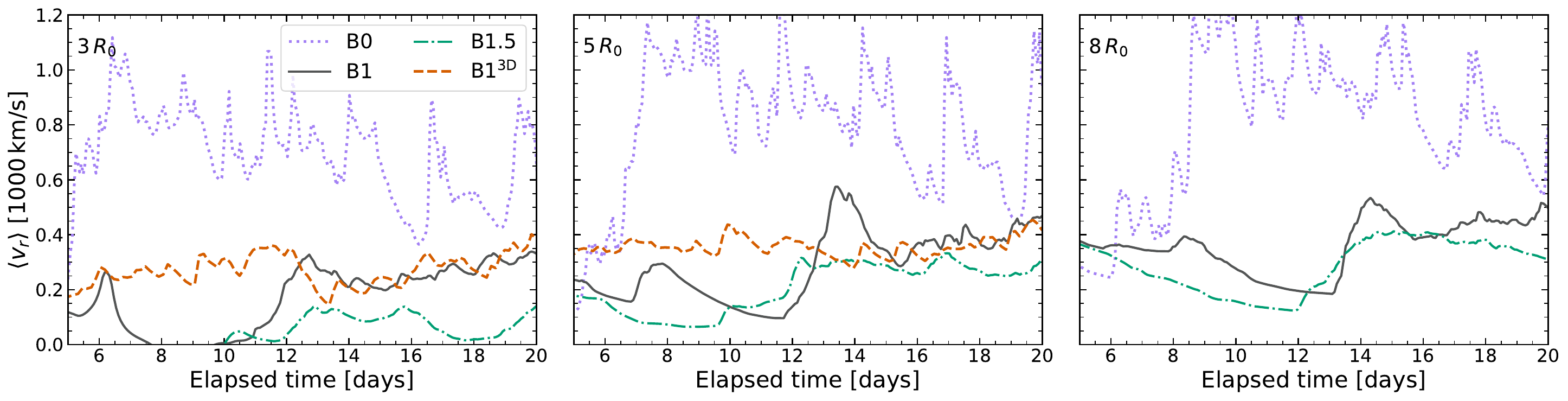}
\caption{Laterally averaged radial velocities for three different radii ($3R_0$, $5R_0$ and $8R_0$) shown as a function of time in days. For the 
B0 model, these radii correspond to approximately $2.0 R_\star$, $3.3 R_\star$ and $5.3 R_\star$, and for the B1 model, these correspond to approximately $1.7 R_\star$, $2.8 R_\star$ and $4.4 R_\star$, and finally for the B1.5 model, these correspond to approximately $1.5 R_\star$, $2.4 R_\star$ and $3.9 R_\star$. For the $\rm B1^{3D}$ model, it corresponds to approximately $1.8 R_\star$ and $3 R_\star$. The colors here represent the different models shown in the legend.}
\label{fig:time_var_v}
\end{figure*}

We further examine how the ionization stratification discussed in Sect.~\ref{ion_stratification} relates to the dynamics of the wind. Fig.~\ref{fig:bsg_snapshot200_probdens} shows the velocity distribution of the B1 model for a single snapshot, with colors showing the ionization fractions of Fe-\textsc{ii}, Fe-\textsc{iii}, and the total line opacity for aNLTE and LTE. As shown in the middle panel, the present treatment qualitatively restructures the iron ionization balance throughout the wind. Compared to the reference LTE model, the present method is strongly recombined to Fe-\textsc{ii}, especially in the slower, denser component. As discussed previously, ionization stratification modifies the line ensemble, clearly visible in the resulting effective line opacities. The comparison between the aNLTE and LTE line opacities shown in the bottom-right panel of Fig.~\ref{fig:bsg_snapshot200_probdens} illustrates this difference in the denser regions, with the $\kappa_{\rm line}$ being relatively higher in such regions due to the restructuring of the ionization species in the present method. Similar to the 2D model, the $\rm B1^{3D}$ model shows similar qualitative radial behavior, with Fe-\textsc{ii} also becoming increasingly important in the outer wind. Due to the smaller radial domain, the region where Fe-\textsc{ii} becomes important is less extended in the $\rm B1^{3D}$ model. 

%\jon{You need to connect this to the more realistic 3D model also.. (just a bit, i think it is similar, yes, but since lower radii we don't reach iron-II, or?}
%JS-NOTE: this we have now checked, so not needed in paper, I'd say.
%We find the models exhibit lateral velocities of $50-70 \rm \, km/s$, which may reduce the effective mass of the star (discussed further in Sect.~\ref{sec:bsg_limitations}). However, the centrifugal mass reduction is only 0.007, 0.015, and 0.017 dex for the B0, B1, and B1.5 models, respectively. Therefore, the new effective mass change is negligible and cannot explain the difference in mass-loss rates between the 2D models and \citet{bjorklund_2023} and LIME.  

%We find for the B0 model a mass-loss rate of approximately $10^{-6} \, \rm  M_\odot/yr$, the B1 model of $3 \times 10^{-6} \, \rm M_\odot/yr$, and the B1.5 model of $3 \times 10^{-6} \, \rm M_\odot/yr$. Similar to the 2D B1 model we find the mass loss rate in the $\rm B1^{3D}$ model to be approximately $3 \times 10^{-6} \, \rm M_\odot/yr$.

\begin{figure*}
\centering
\includegraphics[width=\linewidth]{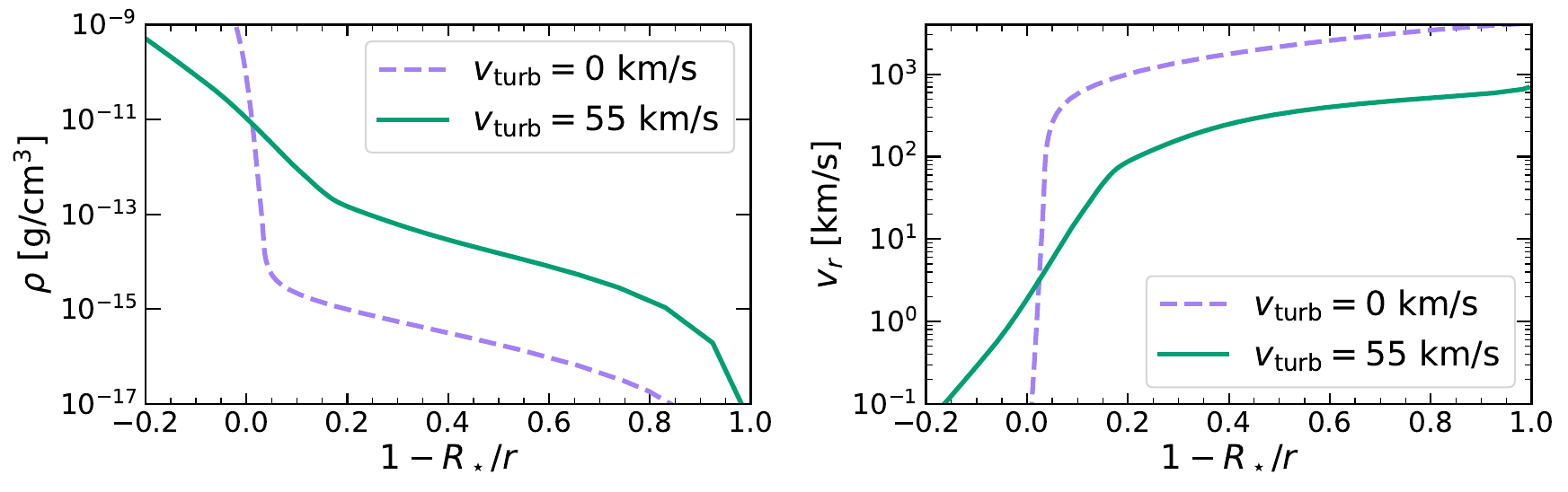}
\caption{Density and radial velocity structure of the hydrodynamically consistent \textsc{FASTWIND} models as a function of the modified radius. The colors indicate the two models, including a $\varv_{\rm turb} = 55 \, \rm km/s$ model and one without any turbulent velocities.}
\label{fig:cmf_rho_v}
\end{figure*}

\subsection{Mass-loss rates}
\label{sec:chapter_bsg_massloss}

Using a few radial cells at the outer edge of the simulation domain, we calculate a temporally and radially averaged mass-loss rate for the models presented in this work, as listed in Table.~\ref{tab:bsg_params}. Comparing the mass-loss rates with the mass-loss fitting formula provided by \citet{bjorklund_2023}, we find a mass-loss rate of $4 \times 10^{-7}$, $3.4 \, \times 10^{-7}$, $ 3 \times 10^{-7}$, and $ 5 \times 10^{-7} \, \rm M_\odot/yr$ for the B0, B1, B1.5 and $\rm B1^{3D}$ model respectively, which are lower than in the 2D and 3D models presented in this work. To match the metallicity adopted in our multi-D simulations, we evaluate the \citet{bjorklund_2023} mass-loss prescription at $Z=0.02$. 

We obtain similar mass loss predictions from \href{https://lime.ster.kuleuven.be/}{LIME}\footnote{https://lime.ster.kuleuven.be/} \citep{sundqvist_lime}, where we find $6.5 \times 10^{-7}, \,  4.2 \times 10^{-7}, \, 4.6 \times 10^{-7}$ and $4.1 \times 10^{-7} \rm \, M_\odot/yr$ for the B0, B1, B1.5 and $\rm B1^{3D}$ model, respectively. In LIME, the continuum contribution to the radiative acceleration is represented using electron scattering, while line acceleration is calculated using a force multiplier formalism under the Sobolev approximation \citep{gayley_1995, luka_2022}. However, around the photosphere of our models, the total continuum contribution is substantially higher than the electron-scattering opacity, resulting in a larger reduction of effective gravity due to continuum opacity sources. 
Accounting for the total continuum opacities, we obtain the photospheric continuum Eddington factors of approximately $0.45$, $0.47$, $0.52$, and $0.52$ for the B0, B1, B1.5, and $\rm B1^{3D}$ models, respectively, significantly exceeding the corresponding electron-scattering Eddington factor, $\Gamma_e$ listed in Table~\ref{tab:bsg_params}. 

To approximately account for this enhanced continuum radiative support in LIME, we increase the luminosity-to-mass ratio so that $\Gamma_e$ in LIME matches the corresponding photospheric total continuum Eddington parameter obtained from the multi-D models. The resulting mass-loss rates increase to approximately $1.2 \times 10^{-6}, \, 8.4 \times 10^{-7}, \, 1.2  \times 10^{-6}$, and $\,10^{-6} \rm \, M_\odot/yr$ for the B0, B1, B1.5 and $\rm B1^{3D}$ models respectively. The revised prediction matches the B0 model, but remains lower by factors of approximately $3.5$, $2.5$ and $3$ for the B1, B1.5 and $\rm B1^{3D}$ models, respectively. Another ingredient that may contribute to such discrepancies is the turbulent pressure associated with substantial velocity fluctuations around the photosphere. This is neither included in the \citet{bjorklund_2023} prescription nor in LIME. We investigate possible implications of such inclusions in Sect.~\ref{sec:1d_comp}.

\subsection{Temporal variations}
\label{sec:temporal_variations}

%Temporal variations in radial wind velocities are much faster in the B0 model, since the wind moves much faster.

%\jon{please insert e.g. Markova and Puls, and potentially other, refs here, since this is well studied.]}
%Due to smaller outward flow velocities, temporal $\langle\varv_r\rangle$ variations in the B1 model also occur over visibly longer timescales than in the B0 model.

In simulations presented here, both regions around the photosphere and the outflow are temporally dynamic. We plot the laterally averaged radial velocities for various radial depths for the different models in Fig.~\ref{fig:time_var_v}. The plot only displays data after the models have relaxed from the initial conditions. The B0 star already has high velocities near the stellar photosphere. Further out, the B0 model on average varies between 600 and 1000 $\rm km/s$. For the B1 model, around the photosphere ($1.5 R_\star \, \text{or} \, 3R_0$ in Fig.\ref{fig:time_var_v}),  shows even more significant $\langle\varv_r\rangle$ temporal variations. The laterally averaged velocity experiences intervals of small outward velocity of a couple of hundred $\rm km/s$, followed by almost negligible average outflow speeds. Further out, wind speeds typically range from $100 - 400  \rm \, km/s$. Qualitatively, the B1.5 model behaves similarly to the B1 model. The 3D model also exhibits substantial temporal variability; at 3 $R_0$ (or 1.8 $R_\star$) the velocity fluctuates approximately between 200 and 400 km/s, whereas at 5 $R_0$ (3 $R_\star$) they typically range between 300 and 450 km/s. Compared to the 2D B1 model, the 3D model shows a more persistent outward flow in the inner wind, without extended time intervals of nearly zero mean radial velocities. Nevertheless, both $\rm B1^{3D}$ and B1, B1.5 models exhibit time-dependent, slowly accelerating outflow compared to the hotter B0 model.

We further investigated temporal variability in the radiative luminosity at $3 R_0$ for the models presented in this work. We constructed a normalized light curve and found the root-mean-square (RMS) flux variations were $9.54\%$, $8.56\%$, $9.85\%$ and $7.79\%$ for the B0, B1, B1.5, and $\rm B1^{3D}$ models, respectively. We also calculated Lomb-Scargle power spectra \citep{Lomb_1976,Scargle_1982} of these luminosity variations and found an excess of power at low temporal frequencies in all our models. Frequencies below which $80 \%$ of the total power remains are approximately 28, 11, 6 and $23 \, \rm \mu Hz$  for the B0, B1, B1.5 and $\rm B1^{3D}$ models, respectively. 
For the 2D set of models, we find that power is concentrated at lower frequencies as we move towards the cooler models. In contrast, the 3D model shows a broader power distribution toward intermediate frequencies than its 2D counterpart. 

\section{Comparison to hydrodynamically consistent 1D models}
\label{sec:1d_comp}

% In the standard formulation, we write the equation of motion as 
% %
% \begin{equation}
% \varv\frac{{\rm d}\varv}{{\rm d}r} = g_{\rm rad}-g
% -\frac{1}{\rho}\frac{{\rm d}P_{\rm gas}}{{\rm d}r},
% \label{eq:cmf_eom_standard}
% \end{equation}
% %
% where, $P_{\rm gas} = \rho a^2$ is the gas pressure, with $a$ as the gas sound speed, $g_{\rm rad}$ is the radiative acceleration, and $g=GM_\star/r^2$ is the gravitational acceleration. \jon{[NOTE: sorry, but are these not introduced before in paper already? at least some?]} 
As discussed in Sect.~\ref{sec:chapter_bsg_massloss}, the mass-loss rates from the 2D and 3D calculations are substantially higher than the ones calculated using the \citet{bjorklund_2023} prescription. However, the underlying models used to create this prescription didn't account for the high turbulent velocities that arise in the multi-D simulations of these stars. Motivated by this, we calculate new 1D, stationary, hydrodynamically consistent wind models following the method introduced by \citet{sundqvist_bjorklund_2019}. These models use the 1D atmosphere code \textsc{FASTWIND} \citep{puls_2020} to solve NLTE radiative transfer in CMF to obtain the radiative acceleration, using the same \textsc{Munich Atomic Database} \citep{pauldrach_1998, puls_2000} as for the Sobolev-based line-force calculations in our multi-D simulations, but we adopt the standard solar metallicity Z=0.013 \citep{asplund_2009}. We iteratively couple the calculated radiative acceleration to the stationary, spherically symmetric equation of motion, allowing us to compute both the mass-loss rate and velocity structure self-consistently. Introducing  turbulent pressure in the equation of motion, $P_{\rm turb}=\rho \varv_{\rm turb}^{2}$ \citep{jiang_2023, Debnath24, nico_2025} the equation of motion is modified as \citep[see also,][]{verhamme_thesis_2025}, 
\begin{equation}
\varv_r\frac{{\rm d}\varv_r}{{\rm d}r} =g_{\rm rad}-g
-\frac{1}{\rho}
\frac{{\rm d}}{{\rm d}r}
\left(P_{\rm gas}+P_{\rm turb}\right).
\label{eq:cmf_eom_turb}
\end{equation}
where, $P_{\rm gas} = \rho a^2$ is the gas pressure, with $a$ as the gas sound speed, $g_{\rm rad}$ is the radiative acceleration, and $g=GM_\star/r^2$ is the gravitational acceleration. We can rewrite the equation using an effective sound speed $a_{\rm eff}^{2}=a^{2}+\varv_{\rm turb}^{2}$ such that \citep[see also,][]{kee_2021},
\begin{equation}
\varv_r\frac{{\rm d}\varv_r}{{\rm d}r}
\left(1-\frac{a_{\rm eff}^{2}}{\varv^{2}}\right)= g_{\rm rad}-g
+\frac{2a_{\rm eff}^{2}}{r}
-\frac{{\rm d}a_{\rm eff}^{2}}{{\rm d}r}.
\label{eq:cmf_eom_turb_expanded}
\end{equation}
Fig.~\ref{fig:cmf_rho_v} shows the density and velocity structure obtained using the method described above for the B1 model parameters with $\varv_{\rm turb} = 55 \, \rm km/s$ and without it. The model with $\varv_{\rm turb} = 55 \, \rm km/s$ shows a much shallower density stratification than the model without any $\varv_{\rm turb}$. The outcome mass-loss rates for the model with $\varv_{\rm turb} = 55 \, \rm km/s$ is $2.87 \times 10^{-6} \, \rm M_\odot/yr$. For the model without any turbulent velocity, the mass-loss rate is $2.22 \times 10^{-7} \, \rm M_\odot/yr$, which is close to the mass-loss rate obtained using the \citet{bjorklund_2023} prescription, with a small difference reflecting the inherent scatter that exists when fitting these models to translate them into an analytic prescription formula and the smaller $Z$ used for this scaling test. Figure~\ref{fig:cmf_rho_v} also demonstrates the difference in the velocity structure, with the unmodified model showcasing a large radial velocity with terminal wind speeds $v_\infty = 4110 \, \rm km/s$, whereas the model including a large turbulent pressure shows a more gradual wind acceleration comparable to the B1 model presented in the paper. The terminal wind for this model is merely $v_\infty = 687 \, \rm km/s$. The global wind properties of this CMF simulation thus resemble the corresponding 2D B1 model well. 

However, computing a CMF model using the average 3D model parameters yields a somewhat different result, as for this model $\varv_{\rm turb} = 25$ km/s. In the CMF models, this results in an in-between case yielding $\dot{M} = 5 \times 10^{-7} \, \rm M_\odot/yr$ and $\varv_\infty = 1900$ km/s, whereas in the 3D simulation we still obtain the higher mass-loss rate and lower wind speeds characteristic of the 2D models with higher turbulent velocities. A likely cause for this discrepancy is the large increase of $\varv_{\rm turb}$ in layers just above the photosphere observed in the 3D model, resulting in values higher than 25 km/s in regions where the line-driven wind is launched, especially when accounting for the fact that the high-density material that carries most of the total wind mass gets accelerated in significantly higher layers than the mean wind \citep{nico_2022b, Debnath24}. However, first tentative attempts to include a radially increasing $\varv_{\rm turb}$ also in the stationary 1D CMF models have not resulted in a significant increase in mass loss (and corresponding lower terminal wind speed), as the radial dependence also induces a counteracting force component due to the turbulent velocity gradient. This difference in sensitivity of the predicted mass-loss rate to the precise value and stratification of $\varv_{\rm turb}$ between the multi-D and 1D CMF models needs to be further investigated in future work.

To further illustrate how turbulent pressure modifies the density stratification, in the approximately quasi-hydrostatic layers below the wind-launching region, we may assume the advective acceleration term to be small. We can then substitute the gas and the turbulent pressure terms in Eq.~\eqref{eq:cmf_eom_turb_expanded} to obtain,
\begin{align}
H_{\rho,\rm turb} &\simeq \frac{a_{\rm eff}^2}{g-g_{\rm rad}} = \frac{a_{\rm eff}^2}{g(1 - \widetilde{\Gamma})} \\
\frac{H_{\rho,\rm turb}}{H_{\rho,0}} &\simeq 1+\frac{\varv_{\rm turb}^2}{a^2}
\label{eq:scaleheight_ratio}
\end{align}
where $H_{\rm \rho, turb}$ is the turbulent velocity-modified density scale height, $H_{\rho,0}$ denotes scale height without turbulent pressure and $\widetilde{\Gamma} = g_{\rm rad}/g$. Because the opacity is higher than electron scattering at the photosphere, we use the total flux-mean opacity \citep[see for definitions][]{luka_2022, nico_2022a} for these calculations. From the above equations, it is clear that as effective gravity decreases, a similar turbulent pressure yields a more geometrically extended atmosphere. For the  B0, B1, B1.5, and $\rm B1^{3D}$ models presented here, we find $H_{\rm \rho, turb}/R_\star$ to be 0.05, 0.08, 0.07, and 0.01, respectively, using turbulent velocity values at the photosphere, compared with $H_{\rm \rho,0}/R_\star$ are 0.0081, 0.0126, 0.0098, and 0.0049. The corresponding density scale height is modified by a factor of approximately 6.9, 6.3, 6.9, and 2.2. Therefore, for models with similar turbulent velocities, effects on the atmospheric density scale height become increasingly important as surface gravity reduces.

%\jon{To understand this... (or something)} 
We consider an atmospheric layer of fixed thickness $\Delta r$ with approximately constant $g-g_{\rm rad}$, $a$, and $\varv_{\rm turb}$. Assuming the same density at the base of the layer, the density enhancement relative to the case without any turbulent velocity can be approximated as
\begin{equation}
\frac{\rho_{\rm c,turb}}{\rho_{\rm c,0}}
\simeq \exp\left[\frac{\Delta r}{H_{\rho,0}} f_{\rm turb}\right],
\label{eq:rho_turb_simple}
\end{equation}
where $f_{\rm turb}=\varv_{\rm turb}^2/a^2_{\rm eff}$. As $\varv_{\rm turb}\gg a$, the factor $f_{\rm turb}$ approaches unity and the density increase becomes less sensitive to further increases in $\varv_{\rm turb}$. At the photosphere, for the 2D B1 model we obtain $f_{\rm turb} = 0.84$, whereas using the 3D values it becomes $0.55$. However, at the effective sonic point, $f_{\rm turb}$ becomes $0.97$ and $0.96$ for the B1 and $\rm B1^{3D}$ models, respectively. Despite the lower photospheric turbulent velocities in the 3D model, the 3D value becomes comparable to the 2D model near their respective effective sonic points. However, this saturation applies only to the simplified fixed-layer density-enhancement approach; in reality, mass losses depend on the radial stratification of the turbulent pressure, radiative acceleration, and the location of the effective sonic point.

\section{Discussion}
\label{sec:discussion}

In this section, we discuss observational implications of the results presented in this work, along with limitations and future improvements, and finally conclude this work. 

\subsection{Observational implications}
\label{sec:observational_imp}

\paragraph{Turbulent velocities and macroturbulence.} The models presented in this work show large velocity dispersions around the photosphere. These turbulent velocities are on the same order as the non-rotational broadening commonly inferred from BSG photospheric absorption lines \citep{Ryans_2002, simon_2017, de_Burgos_2024, nadya_2024}. For a 3D O star, \citet{lara25} showed that broad photospheric spectral lines can naturally emerge without ad hoc extra broadening terms. Turbulent velocities found here will likely correspond to characteristic non-rotational line broadening of order several tens of ${\rm km/s}$, comparable to observed trends. Future 3D spectral synthesis calculations of these stars are required to confirm whether this photospheric line-broadening due to subsurface turbulence seen in O stars extends towards the BSG regime. 
Further, BSGs have been observed to display significant wind line-profile variability \citep{Prinja_2002, markova_2008, haucke_2018}. As such, the temporal variations seen in this work may indeed point towards give rise to line-profile variability. Our models also suggest 
that velocity inferred from spectral diagnostics may sample different components in the outflow and therefore vary between different observed epochs. 

%\jon{NOTE: general comment. Here, and at a few other places, you sort of mix in discussion of observational results directly with the results of simulations. It would be more consustent to lift this over to the discussion section, since you do have such section there anyway. That would make paper more readable, in my opinion.]}

% We further calculate the Lomb-Scargle power spectra of these normalized luminosity variations. We found that all three models display an excess of power at low frequencies, characteristic of stochastic low-frequency (SLF) variability. Additionally, frequencies below which $80 \%$ of the total power remains are approximately, $28 \, \rm \mu Hz$, $11 \, \rm \mu Hz$, and $6 \, \rm \mu Hz$ for the B0, B1 and B1.5 models respectively. For comparison, \citet{Ma_2026} finds the $80 \%$ power threshold to be approximately below $20 \, \rm \mu Hz$ in their synthetic BSG light curves. This shift towards lower characteristic frequencies from the B0 to B1.5 model is consistent with the longer temporal variability timescales apparent in the radial velocity variations. \jon{[NOTE: similar here, you mix your results, with discussion of other results..]}  

\paragraph{Mass-loss rates and spectroscopic masses.}  Recent empirical studies of the Galactic, LMC, and SMC BSGs do not find such high mass-loss rates  \citep{de_Burgos_2024, verhamme_2024, matheus_2025, verhamme_2026}; rather, the empirically found mass losses overall align with the unmodified \citet{bjorklund_2023} prescription. 
For determining the inferred spectroscopic masses for these stars, the turbulent pressure support predicted by multi-D simulations was not included. Previous comparisons between 1D and multi-D O stars have shown that including the turbulent pressure may change the spectroscopically inferred surface gravity, leading to a larger spectroscopic mass \citep{gonzalez_2025, verhamme_thesis_2025}. In a case where such trends continue towards the BSG regimes, spectroscopic masses derived without turbulent pressure could be systematically underestimated. As such, an upward revision of spectroscopic mass would correspond to a lower inferred classical Eddington $\Gamma_e$. As a consequence, the observed BSGs in the empirical studies may in reality occupy a somewhat lower luminosity-to-mass once turbulent pressure is treated consistently in the empirical studies.    

\paragraph{Stochastic low-frequency variability.} A direct observational consequence of the models presented here is the strong temporal variability. As shown in Sect.~\ref{sec:temporal_variations}, all our models exhibit stochastic luminosity fluctuations with an excess of power at lower frequencies. These amplitude variations are of a similar order to the bolometric luminosity variability of $\sim 10\%$ found in the recent 3D BSG envelope simulations by \citet{Ma_2026}. However, the simulations presented here are unified atmosphere and wind models, whereas \citet{Ma_2026} considers a 3D envelope model without a line-driven wind outflow. Additionally, \citet{Ma_2026} finds the $80 \%$ power threshold to be approximately below $20 \, \rm \mu Hz$ in their synthetic BSG light curves. As shown in Sect.~\ref{sec:temporal_variations}, we find a similar excess of power at low frequencies. Such SLF variability has been detected in BSGs using high-precision space-based photometry \citep{Bowman2019, Ma_2024, Kourniotis_2025}. The physical origin of the observed SLF is still debated. Several mechanisms have been proposed, such as internal gravity waves generated by core convection \citep{Bowman2019b, Bowman2020} and turbulence associated with the iron-opacity bump \citep{schultz_2022}. The calculations presented in this work indicate that for BSGs in the stellar parameter range considered here, turbulence generated in the subsurface iron-opacity bump may at least qualitatively explain the observed SLF variability \citep[see also,][]{Ma_2026}. However, since we do not model the stellar core, we can't rule out an internal gravity wave contribution to the SLF variability. Recent 3D hydrodynamical simulations by \citet{Pathak_2026}, which include both core and envelope convective regions, suggest that simulations with only the outer envelope convection zone produce less power at low frequencies compared to their full star configuration. Work is currently underway to compute observable signatures from such simulations to compare with high-resolution spectroscopic observations \citep[e.g.,][]{ankur_2025}.

%\jon{ try to see if you can cite some of Praneet's work here perhaps? A paper, and one "in prep" perhaps? (good for further collaborations with them, Dom, etc..)} 

\subsection{Limitations and next steps}
\label{sec:bsg_limitations}

% \item First, the BSG models presented here show lateral velocity flows in their extended atmosphere. Such lateral velocities effectively spin the star with modest rotational velocities of $\sim 50-70 \rm  \, km/s$. However, the reduction in the effective mass is negligible, and so is the influence on the mass-loss rates we predict. Nonetheless, further investigation is required to understand these lateral motions. \jon{Take this first oneout. We will deal with it, yes, but no need to mention in paper, if you ask me.. unnecessary distraction.}

The calculations presented in this chapter demonstrate that the aNLTE treatment can substantially modify the ionization balance in the wind, influencing the line-force parameters and affecting the dynamical properties of BSG winds. However, a few aspects of the BSG models presented in this chapter need deeper understanding and further investigation.

\begin{itemize}
    \item First, the current models in this chapter are limited to a small subset of the BSG parameter space. A natural extension would be to construct a larger set of 3D models covering a broader range of temperature (extending towards the colder side of bistability), surface gravity, luminosities, and metallicity. Such a grid would allow us to make systematic predictions of how the ionization behavior influences the dynamics of these stars. 
    \item Second, in the FLD closure relation, the flux is determined by the gradient of the radiation energy density, and lateral gradients in the energy density may generate a lateral radiative flux and radiation force \citep{nico_2022a}. Since FLD constrains the flux direction to follow the gradient of the radiation energy density, the approximation becomes less reliable as the radiation field becomes more anisotropic \citep{moens_thesis_2023}, which is likely the case around the photosphere. Work is currently underway to replace the FLD with a two-moment radiation transport scheme, M1, for the massive RHD simulations, which evolves both the radiation energy density and the flux and does not involve constraining the flux from the gradient of the energy density. 
    \item Third, the pseudo-planar geometry adopted in these simulations. As the lateral coordinate remains locally Cartesian, the pseudo-planar geometry can be regarded only as a local patch of the stellar atmosphere. This approximation becomes less applicable when the lateral structures are comparatively large. 
    \item Lastly, the present calculations employ a diluted frequency-integrated radiation field and, thereby, integrated mean opacities. Although this provides an efficient way to incorporate aNLTE effects in multi-D calculations, the ionization balance is intrinsically a frequency-dependent problem. The BSGs have a rich spectrum, and using an integrated radiation field may over- or underestimate how different parts of the spectrum interact with the gas. Work is currently underway to develop a frequency-dependent NLTE opacity formalism including line-driving for multi-D RHD simulations. 
\end{itemize}

%Since 2D affects the numerical values of the turbulent velocities, a natural extension of these models to 3D is required to understand these turbulent pressure-induced mass-loss rates. As discussed, in these low surface-gravity regimes, global properties such as mass-loss rates may be affected by turbulence. For a robust prediction of mass-loss rates, the framework thus needs to be benchmarked in 3D. 
 
\subsection{Summary and Conclusion}
\label{sec:bsg_conc}

In this work, we presented the first unified multi-D atmosphere and wind models of early BSGs. We included aNLTE effects in the ionization balance and in calculating the line opacities. We conducted our study with three 2D and one 3D model spanning $T_{\rm eff}=22.7$--$30.4,{\rm kK}$. Similar to the multi-D O stars \citep{jiang_2015, Debnath24, cassie_25, lara25, nico_2025}, we find large velocity fluctuations originating deep within the stellar envelope, which propagate to the optical photosphere. Compared to the 2D model, the 3D model shows reduced photospheric turbulent velocities, lower atmospheric inflation, and less pronounced fall-back of the mean radial velocity at the photosphere. These differences suggest that while 2D models capture qualitatively the mechanisms governing the atmosphere and wind, 3D simulations are necessary for quantitative characterization of the atmospheric dynamics. Due to their low surface gravities, turbulent pressure modifies the density scale height around the photosphere to such an extent that it influences the global mass-loss rates. As a result, for the same parameter, we find higher mass-loss rates than the \citet{bjorklund_2023} prescriptions, which are based on 1D hydrodynamically consistent \textsc{FASTWIND} models \citep{sundqvist_bjorklund_2019}. When computed using turbulent pressure obtained from multi-D, the mass-loss rates from 1D hydrodynamically consistent \textsc{FASTWIND} are higher, and the terminal wind speeds are lower. Further, we found that in our aNLTE treatment, the ionization stratification in the wind significantly differs from the reference LTE ionization. Moreover, the wind dynamics also differ between the models, with the hotter B0 model exhibiting a faster wind and the cooler B1 and B1.5 displaying a gradual and slower wind. Finally, we also found significant temporal variations in our models; the photospheric radiative luminosity varies by an order of $10\%$. The simulations also show an excess of power at low temporal frequencies similar to observed trends.

\begin{acknowledgements}
    The computational resources used for this work were provided by Vlaams Supercomputer Centrum (VSC), funded by the Research Foundation-Flanders (FWO) and the Flemish Government. DD, JS, and NM acknowledge support from the European Research Council (ERC) under the Horizon Europe grant agreement 101044048. JS further acknowledges support from the Belgian Research Foundation Flanders (FWO) Odysseus program under grant number G0H9218N, from FWO grant G077822N, and from KU Leuven C1 grant BRAVE C16/23/009. The authors also acknowledge the fruitful discussions, suggestions, and comments from the present and past members of the EQUATION team. Data analysis and visualisation in this project relied on WindSimTools \citep{nico_2025}, NumPy \citep{harris_2020}, SciPy \citep{virtanen_2020}, and Matplotlib \citep{hunter_2007}.

\end{acknowledgements}

\bibliographystyle{aa}
\bibliography{references_ostar}

@ARTICLE{asplund_2009,
       author = {{Asplund}, Martin and {Grevesse}, Nicolas and {Sauval}, A. Jacques and {Scott}, Pat},
        title = "{The Chemical Composition of the Sun}",
      journal = {\araa},
         year = 2009,
        month = sep,
       volume = {47},
       number = {1},
        pages = {481-522},
          doi = {10.1146/annurev.astro.46.060407.145222},
archivePrefix = {arXiv},
       eprint = {0909.0948},
 primaryClass = {astro-ph.SR},
       adsurl = {https://ui.adsabs.harvard.edu/abs/2009ARA&A..47..481A}
}

@ARTICLE{cantiello_2009,
       author = {{Cantiello}, M. and {Langer}, N. and {Brott}, I. and {de Koter}, A. and {Shore}, S.~N. and {Vink}, J.~S. and {Voegler}, A. and {Lennon}, D.~J. and {Yoon}, S. -C.},
        title = "{Sub-surface convection zones in hot massive stars and their observable consequences}",
      journal = {\aap},
         year = 2009,
        month = may,
       volume = {499},
       number = {1},
        pages = {279-290},
          doi = {10.1051/0004-6361/200911643},
archivePrefix = {arXiv},
       eprint = {0903.2049},
 primaryClass = {astro-ph.SR},
       adsurl = {https://ui.adsabs.harvard.edu/abs/2009A&A...499..279C}
}

@ARTICLE{cak1975,
       author = {{Castor}, J.~I. and {Abbott}, D.~C. and {Klein}, R.~I.},
        title = "{Radiation-driven winds in Of stars.}",
      journal = {\apj},
         year = 1975,
        month = jan,
       volume = {195},
        pages = {157-174},
          doi = {10.1086/153315},
       adsurl = {https://ui.adsabs.harvard.edu/abs/1975ApJ...195..157C}
}

@ARTICLE{gayley_1995,
       author = {{Gayley}, Kenneth G.},
        title = "{An Improved Line-Strength Parameterization in Hot-Star Winds}",
      journal = {\apj},
         year = 1995,
        month = nov,
       volume = {454},
        pages = {410},
          doi = {10.1086/176492},
       adsurl = {https://ui.adsabs.harvard.edu/abs/1995ApJ...454..410G}
}

@ARTICLE{harris_2020,
       author = {{Harris}, Charles R. and {Millman}, K. Jarrod and {van der Walt}, St{\'e}fan J. and {Gommers}, Ralf and {Virtanen}, Pauli and {Cournapeau}, David and {Wieser}, Eric and {Taylor}, Julian and {Berg}, Sebastian and {Smith}, Nathaniel J. and {Kern}, Robert and {Picus}, Matti and {Hoyer}, Stephan and {van Kerkwijk}, Marten H. and {Brett}, Matthew and {Haldane}, Allan and {del R{\'\i}o}, Jaime Fern{\'a}ndez and {Wiebe}, Mark and {Peterson}, Pearu and {G{\'e}rard-Marchant}, Pierre and {Sheppard}, Kevin and {Reddy}, Tyler and {Weckesser}, Warren and {Abbasi}, Hameer and {Gohlke}, Christoph and {Oliphant}, Travis E.},
        title = "{Array programming with NumPy}",
      journal = {\nat},
         year = 2020,
        month = sep,
       volume = {585},
       number = {7825},
        pages = {357-362},
          doi = {10.1038/s41586-020-2649-2},
archivePrefix = {arXiv},
       eprint = {2006.10256},
 primaryClass = {cs.MS},
       adsurl = {https://ui.adsabs.harvard.edu/abs/2020Natur.585..357H}
}

@ARTICLE{hawcroft_2023,
       author = {{Hawcroft}, C. and {Sana}, H. and {Mahy}, L. and {Sundqvist}, J.~O. and {de Koter}, A. and {Crowther}, P.~A. and {Bestenlehner}, J.~M. and {Brands}, S.~A. and {David-Uraz}, A. and {Decin}, L. and {Erba}, C. and {Garcia}, M. and {Hamann}, W. -R. and {Herrero}, A. and {Ignace}, R. and {Kee}, N.~D. and {Kub{\'a}tov{\'a}}, B. and {Lefever}, R. and {Moffat}, A. and {Najarro}, F. and {Oskinova}, L. and {Pauli}, D. and {Prinja}, R. and {Puls}, J. and {Sander}, A.~A.~C. and {Shenar}, T. and {St-Louis}, N. and {ud-Doula}, A. and {Vink}, J.~S.},
        title = "{X-Shooting ULLYSES: Massive stars at low metallicity. III. Terminal wind speeds of ULLYSES massive stars}",
      journal = {arXiv e-prints},
         year = 2023,
        month = mar,
          eid = {arXiv:2303.12165},
        pages = {arXiv:2303.12165},
          doi = {10.48550/arXiv.2303.12165},
archivePrefix = {arXiv},
       eprint = {2303.12165},
 primaryClass = {astro-ph.SR},
       adsurl = {https://ui.adsabs.harvard.edu/abs/2023arXiv230312165H}
}

@ARTICLE{hunter_2007,
       author = {{Hunter}, John D.},
        title = "{Matplotlib: A 2D Graphics Environment}",
      journal = {Computing in Science and Engineering},
         year = 2007,
        month = may,
       volume = {9},
       number = {3},
        pages = {90-95},
          doi = {10.1109/MCSE.2007.55},
       adsurl = {https://ui.adsabs.harvard.edu/abs/2007CSE.....9...90H}
}

@ARTICLE{opal,
       author = {{Iglesias}, Carlos A. and {Rogers}, Forrest J.},
        title = "{Updated Opal Opacities}",
      journal = {\apj},
         year = 1996,
        month = jun,
       volume = {464},
        pages = {943},
          doi = {10.1086/177381},
       adsurl = {https://ui.adsabs.harvard.edu/abs/1996ApJ...464..943I}
}

@ARTICLE{jiang_2015,
       author = {{Jiang}, Yan-Fei and {Cantiello}, Matteo and {Bildsten}, Lars and {Quataert}, Eliot and {Blaes}, Omer},
        title = "{Local Radiation Hydrodynamic Simulations of Massive Star Envelopes at the Iron Opacity Peak}",
      journal = {\apj},
         year = 2015,
        month = nov,
       volume = {813},
       number = {1},
          eid = {74},
        pages = {74},
          doi = {10.1088/0004-637X/813/1/74},
archivePrefix = {arXiv},
       eprint = {1509.05417},
 primaryClass = {astro-ph.SR},
       adsurl = {https://ui.adsabs.harvard.edu/abs/2015ApJ...813...74J}
}

@ARTICLE{keppens_2020,
       author = {{Keppens}, Rony and {Teunissen}, Jannis and {Xia}, Chun and {Porth}, Oliver},
        title = "{MPI-AMRVAC: a parallel, grid-adaptive PDE toolkit}",
      journal = {arXiv e-prints},
         year = 2020,
        month = apr,
          eid = {arXiv:2004.03275},
        pages = {arXiv:2004.03275},
          doi = {10.48550/arXiv.2004.03275},
archivePrefix = {arXiv},
       eprint = {2004.03275},
 primaryClass = {astro-ph.IM},
       adsurl = {https://ui.adsabs.harvard.edu/abs/2020arXiv200403275K}
}

@ARTICLE{lucy_1993,
       author = {{Lucy}, L.~B. and {Abbott}, D.~C.},
        title = "{Multiline Transfer and the Dynamics of Wolf-Rayet Winds}",
      journal = {\apj},
         year = 1993,
        month = mar,
       volume = {405},
        pages = {738},
          doi = {10.1086/172402},
       adsurl = {https://ui.adsabs.harvard.edu/abs/1993ApJ...405..738L}
}

@ARTICLE{nico_2022b,
       author = {{Moens}, N. and {Poniatowski}, L.~G. and {Hennicker}, L. and {Sundqvist}, J.~O. and {El Mellah}, I. and {Kee}, N.~D.},
        title = "{First 3D radiation-hydrodynamic simulations of Wolf-Rayet winds}",
      journal = {\aap},
         year = {2022b},
        month = sep,
       volume = {665},
          eid = {A42},
        pages = {A42},
          doi = {10.1051/0004-6361/202243451},
archivePrefix = {arXiv},
       eprint = {2203.01108},
 primaryClass = {astro-ph.SR},
       adsurl = {https://ui.adsabs.harvard.edu/abs/2022A&A...665A..42M},
      sortkey = {02}
}

@ARTICLE{nico_2022a,
       author = {{Moens}, N. and {Sundqvist}, J.~O. and {El Mellah}, I. and {Poniatowski}, L. and {Teunissen}, J. and {Keppens}, R.},
        title = "{Radiation-hydrodynamics with MPI-AMRVAC . Flux-limited diffusion}",
      journal = {\aap},
         year = {2022a},
        month = jan,
       volume = {657},
          eid = {A81},
        pages = {A81},
          doi = {10.1051/0004-6361/202141023},
archivePrefix = {arXiv},
       eprint = {2104.03968},
 primaryClass = {astro-ph.IM},
       adsurl = {https://ui.adsabs.harvard.edu/abs/2022A&A...657A..81M},
      sortkey = {01}
}

@ARTICLE{pauldrach_1986,
       author = {{Pauldrach}, A. and {Puls}, J. and {Kudritzki}, R.~P.},
        title = "{Radiation-driven winds of hot luminous stars. Improvements of the theory and first results.}",
      journal = {\aap},
         year = 1986,
        month = aug,
       volume = {164},
        pages = {86-100},
       adsurl = {https://ui.adsabs.harvard.edu/abs/1986A&A...164...86P}
}

@INPROCEEDINGS{pauldrach_1998,
       author = {{Pauldrach}, A.~W.~A. and {Lennon}, M. and {Hoffmann}, T.~L. and {Sellmaier}, F. and {Kudritzki}, R. -P. and {Puls}, J.},
        title = "{Realistic Models for Expanding Atmospheres}",
    booktitle = {Properties of Hot Luminous Stars},
         year = 1998,
       editor = {{Howarth}, Ian},
       series = {Astronomical Society of the Pacific Conference Series},
       volume = {131},
        month = jan,
        pages = {258},
       adsurl = {https://ui.adsabs.harvard.edu/abs/1998ASPC..131..258P}
}

@ARTICLE{luka_2021,
       author = {{Poniatowski}, L.~G. and {Sundqvist}, J.~O. and {Kee}, N.~D. and {Owocki}, S.~P. and {Marchant}, P. and {Decin}, L. and {de Koter}, A. and {Mahy}, L. and {Sana}, H.},
        title = "{Dynamically inflated wind models of classical Wolf-Rayet stars}",
      journal = {\aap},
         year = 2021,
        month = mar,
       volume = {647},
          eid = {A151},
        pages = {A151},
          doi = {10.1051/0004-6361/202039595},
archivePrefix = {arXiv},
       eprint = {2012.05823},
 primaryClass = {astro-ph.SR},
       adsurl = {https://ui.adsabs.harvard.edu/abs/2021A&A...647A.151P}
}

@ARTICLE{luka_2022,
       author = {{Poniatowski}, L.~G. and {Kee}, N.~D. and {Sundqvist}, J.~O. and {Driessen}, F.~A. and {Moens}, N. and {Owocki}, S.~P. and {Gayley}, K.~G. and {Decin}, L. and {de Koter}, A. and {Sana}, H.},
        title = "{Method and new tabulations for flux-weighted line opacity and radiation line force in supersonic media}",
      journal = {\aap},
         year = 2022,
        month = nov,
       volume = {667},
          eid = {A113},
        pages = {A113},
          doi = {10.1051/0004-6361/202142888},
archivePrefix = {arXiv},
       eprint = {2204.09981},
 primaryClass = {astro-ph.SR},
       adsurl = {https://ui.adsabs.harvard.edu/abs/2022A&A...667A.113P}
}

@ARTICLE{puls_2000,
       author = {{Puls}, J. and {Springmann}, U. and {Lennon}, M.},
        title = "{Radiation driven winds of hot luminous stars. XIV. Line statistics and radiative driving}",
      journal = {\aaps},
         year = 2000,
        month = jan,
       volume = {141},
        pages = {23-64},
          doi = {10.1051/aas:2000312},
       adsurl = {https://ui.adsabs.harvard.edu/abs/2000A&AS..141...23P}
}

@ARTICLE{iglesias_1992,
       author = {{Rogers}, Forrest J. and {Iglesias}, Carlos A.},
        title = "{Radiative Atomic Rosseland Mean Opacity Tables}",
      journal = {\apjs},
         year = 1992,
        month = apr,
       volume = {79},
        pages = {507},
          doi = {10.1086/191659},
       adsurl = {https://ui.adsabs.harvard.edu/abs/1992ApJS...79..507R}
}

@article{schultz_2022,
doi = {10.3847/2041-8213/ac441f},
url = {https://dx.doi.org/10.3847/2041-8213/ac441f},
year = {2022},
month = {jan},
publisher = {The American Astronomical Society},
volume = {924},
number = {1},
pages = {L11},
author = {William C. Schultz and Lars Bildsten and Yan-Fei Jiang},
title = {Stochastic Low-frequency Variability in Three-dimensional Radiation Hydrodynamical Models of Massive Star Envelopes},
journal = {The Astrophysical Journal Letters}
}

@BOOK{schwarzschild_1958,
       author = {{Schwarzschild}, Martin},
        title = "{Structure and evolution of the stars.}",
         year = 1958,
       adsurl = {https://ui.adsabs.harvard.edu/abs/1958ses..book.....S}
}

@ARTICLE{simon_2017,
       author = {{Sim{\'o}n-D{\'\i}az}, S. and {Godart}, M. and {Castro}, N. and {Herrero}, A. and {Aerts}, C. and {Puls}, J. and {Telting}, J. and {Grassitelli}, L.},
        title = "{The IACOB project . III. New observational clues to understand macroturbulent broadening in massive O- and B-type stars}",
      journal = {\aap},
         year = 2017,
        month = jan,
       volume = {597},
          eid = {A22},
        pages = {A22},
          doi = {10.1051/0004-6361/201628541},
archivePrefix = {arXiv},
       eprint = {1608.05508},
 primaryClass = {astro-ph.SR},
       adsurl = {https://ui.adsabs.harvard.edu/abs/2017A&A...597A..22S}
}

@BOOK{sobolev_1960,
       author = {{Sobolev}, V.~V.},
        title = "{Moving Envelopes of Stars}",
         year = 1960,
          doi = {10.4159/harvard.9780674864658},
       adsurl = {https://ui.adsabs.harvard.edu/abs/1960mes..book.....S}
}

@ARTICLE{sundqvist_2018,
       author = {{Sundqvist}, J.~O. and {Owocki}, S.~P. and {Puls}, J.},
        title = "{2D wind clumping in hot, massive stars from hydrodynamical line-driven instability simulations using a pseudo-planar approach}",
      journal = {\aap},
         year = 2018,
        month = mar,
       volume = {611},
          eid = {A17},
        pages = {A17},
          doi = {10.1051/0004-6361/201731718},
archivePrefix = {arXiv},
       eprint = {1710.07780},
 primaryClass = {astro-ph.SR},
       adsurl = {https://ui.adsabs.harvard.edu/abs/2018A&A...611A..17S}
}

@ARTICLE{sundqvist_bjorklund_2019,
       author = {{Sundqvist}, J.~O. and {Bj{\"o}rklund}, R. and {Puls}, J. and {Najarro}, F.},
        title = "{New predictions for radiation-driven, steady-state mass-loss and wind-momentum from hot, massive stars. I. Method and first results}",
      journal = {\aap},
         year = 2019,
        month = dec,
       volume = {632},
          eid = {A126},
        pages = {A126},
          doi = {10.1051/0004-6361/201936580},
archivePrefix = {arXiv},
       eprint = {1910.06586},
 primaryClass = {astro-ph.SR},
       adsurl = {https://ui.adsabs.harvard.edu/abs/2019A&A...632A.126S}
}

@ARTICLE{vink_2001,
       author = {{Vink}, Jorick S. and {de Koter}, A. and {Lamers}, H.~J.~G.~L.~M.},
        title = "{Mass-loss predictions for O and B stars as a function of metallicity}",
      journal = {\aap},
         year = 2001,
        month = apr,
       volume = {369},
        pages = {574-588},
          doi = {10.1051/0004-6361:20010127},
archivePrefix = {arXiv},
       eprint = {astro-ph/0101509},
 primaryClass = {astro-ph},
       adsurl = {https://ui.adsabs.harvard.edu/abs/2001A&A...369..574V}
}

@ARTICLE{virtanen_2020,
       author = {{Virtanen}, Pauli and {Gommers}, Ralf and {Oliphant}, Travis E. and {Haberland}, Matt and {Reddy}, Tyler and {Cournapeau}, David and {Burovski}, Evgeni and {Peterson}, Pearu and {Weckesser}, Warren and {Bright}, Jonathan and {van der Walt}, St{\'e}fan J. and {Brett}, Matthew and {Wilson}, Joshua and {Millman}, K. Jarrod and {Mayorov}, Nikolay and {Nelson}, Andrew R.~J. and {Jones}, Eric and {Kern}, Robert and {Larson}, Eric and {Carey}, C.~J. and {Polat}, {\.I}lhan and {Feng}, Yu and {Moore}, Eric W. and {VanderPlas}, Jake and {Laxalde}, Denis and {Perktold}, Josef and {Cimrman}, Robert and {Henriksen}, Ian and {Quintero}, E.~A. and {Harris}, Charles R. and {Archibald}, Anne M. and {Ribeiro}, Ant{\^o}nio H. and {Pedregosa}, Fabian and {van Mulbregt}, Paul and {SciPy 1. 0 Contributors}},
        title = "{SciPy 1.0: fundamental algorithms for scientific computing in Python}",
      journal = {Nature Methods},
         year = 2020,
        month = feb,
       volume = {17},
        pages = {261-272},
          doi = {10.1038/s41592-019-0686-2},
archivePrefix = {arXiv},
       eprint = {1907.10121},
 primaryClass = {cs.MS},
       adsurl = {https://ui.adsabs.harvard.edu/abs/2020NatMe..17..261V}
}

@article{xia_2018,
doi = {10.3847/1538-4365/aaa6c8},
url = {https://dx.doi.org/10.3847/1538-4365/aaa6c8},
year = {2018},
month = {feb},
publisher = {The American Astronomical Society},
volume = {234},
number = {2},
pages = {30},
author = {C. Xia and J. Teunissen and I. El Mellah and E. Chané and R. Keppens},
title = {MPI-AMRVAC 2.0 for Solar and Astrophysical Applications},
journal = {The Astrophysical Journal Supplement Series}
}

@ARTICLE{jiang_2023,
       author = {{Jiang}, Yan-Fei},
        title = "{Three Dimensional Natures of Massive Star Envelopes}",
      journal = {Galaxies},
         year = 2023,
        month = oct,
       volume = {11},
       number = {5},
          eid = {105},
        pages = {105},
          doi = {10.3390/galaxies11050105},
archivePrefix = {arXiv},
       eprint = {2310.07829},
 primaryClass = {astro-ph.SR},
       adsurl = {https://ui.adsabs.harvard.edu/abs/2023Galax..11..105J}
}

@ARTICLE{schultz_2020,
       author = {{Schultz}, William C. and {Bildsten}, Lars and {Jiang}, Yan-Fei},
        title = "{Convectively Driven 3D Turbulence in Massive Star Envelopes. I. A 1D Implementation of Diffusive Radiative Transport}",
      journal = {\apj},
         year = 2020,
        month = oct,
       volume = {902},
       number = {1},
          eid = {67},
        pages = {67},
          doi = {10.3847/1538-4357/abb405},
archivePrefix = {arXiv},
       eprint = {2009.01238},
 primaryClass = {astro-ph.SR},
       adsurl = {https://ui.adsabs.harvard.edu/abs/2020ApJ...902...67S}
}

@INPROCEEDINGS{petrenz_puls_2000,
       author = {{Petrenz}, P. and {Puls}, J.},
        title = "{Selfconsistent Models of Winds from Rotating Early-Type Stars - Application to B[e] star winds}",
    booktitle = {IAU Colloq. 175: The Be Phenomenon in Early-Type Stars},
         year = 2000,
       editor = {{Smith}, Myron A. and {Henrichs}, Huib F. and {Fabregat}, Juan},
       series = {Astronomical Society of the Pacific Conference Series},
       volume = {214},
        month = jan,
        pages = {626},
       adsurl = {https://ui.adsabs.harvard.edu/abs/2000ASPC..214..626P}
}

@ARTICLE{Debnath24,
       author = {{Debnath}, D. and {Sundqvist}, J.~O. and {Moens}, N. and {Van der Sijpt}, C. and {Verhamme}, O. and {Poniatowski}, L.~G.},
        title = "{2D unified atmosphere and wind simulations of O-type stars}",
      journal = {\aap},
         year = 2024,
        month = apr,
       volume = {684},
          eid = {A177},
        pages = {A177},
          doi = {10.1051/0004-6361/202348206},
archivePrefix = {arXiv},
       eprint = {2401.08391},
 primaryClass = {astro-ph.SR},
       adsurl = {https://ui.adsabs.harvard.edu/abs/2024A&A...684A.177D}
}

@ARTICLE{cassie_25,
       author = {{Van der Sijpt}, C. and {Sundqvist}, J.~O. and {Debnath}, D. and {Driessen}, F.~A. and {Moens}, N.},
        title = "{Structure formation in O-type stars and Wolf{\textendash}Rayet stars}",
      journal = {\aap},
         year = 2025,
        month = feb,
       volume = {694},
          eid = {A54},
        pages = {A54},
          doi = {10.1051/0004-6361/202452385},
archivePrefix = {arXiv},
       eprint = {2501.09378},
 primaryClass = {astro-ph.SR},
       adsurl = {https://ui.adsabs.harvard.edu/abs/2025A&A...694A..54V}
}

@ARTICLE{lara25,
       author = {{Delbroek}, L. and {Sundqvist}, J.~O. and {Debnath}, D. and {Moens}, N. and {Backs}, F. and {Van der Sijpt}, C. and {Verhamme}, O. and {Schillemans}, P.},
        title = "{Spectral synthesis of 3D unified model atmospheres with winds for O stars}",
      journal = {\aap},
         year = 2025,
        month = dec,
       volume = {704},
          eid = {A104},
        pages = {A104},
          doi = {10.1051/0004-6361/202556421},
archivePrefix = {arXiv},
       eprint = {2507.12057},
 primaryClass = {astro-ph.SR},
       adsurl = {https://ui.adsabs.harvard.edu/abs/2025A&A...704A.104D}
}

@ARTICLE{sundqvist_lime,
       author = {{Sundqvist}, J.~O. and {Debnath}, D. and {Backs}, F. and {Verhamme}, O. and {Moens}, N. and {Delbroek}, L. and {Dickson}, D. and {Schillemans}, P. and {Van der Sijpt}, C. and {Dirickx}, M.},
        title = "{An on-the-fly line-driven-wind iterative mass-loss estimator (LIME) for hot, massive stars of arbitrary chemical compositions}",
      journal = {arXiv e-prints},
         year = 2025,
        month = apr,
          eid = {arXiv:2504.00936},
        pages = {arXiv:2504.00936},
          doi = {10.48550/arXiv.2504.00936},
archivePrefix = {arXiv},
       eprint = {2504.00936},
 primaryClass = {astro-ph.SR},
       adsurl = {https://ui.adsabs.harvard.edu/abs/2025arXiv250400936S}
}

@ARTICLE{verhamme_2024,
       author = {{Verhamme}, O. and {Sundqvist}, J. and {de Koter}, A. and {Sana}, H. and {Backs}, F. and {Brands}, S.~A. and {Najarro}, F. and {Puls}, J. and {Vink}, J.~S. and {Crowther}, P.~A. and {Kub{\'a}tov{\'a}}, B. and {Sander}, A.~A.~C. and {Bernini-Peron}, M. and {Kuiper}, R. and {Prinja}, R.~K. and {Schillemans}, P. and {Shenar}, T. and {van Loon}, J. Th. and {XShootu collaboration}},
        title = "{X-Shooting ULLYSES: Massive Stars at low metallicity: IX. Empirical constraints on mass-loss rates and clumping parameters for OB supergiants in the Large Magellanic Cloud}",
      journal = {\aap},
         year = 2024,
        month = dec,
       volume = {692},
          eid = {A91},
        pages = {A91},
          doi = {10.1051/0004-6361/202451169},
archivePrefix = {arXiv},
       eprint = {2410.14937},
 primaryClass = {astro-ph.SR},
       adsurl = {https://ui.adsabs.harvard.edu/abs/2024A&A...692A..91V}
}

@ARTICLE{nico_2025,
       author = {{Moens}, N. and {Debnath}, D. and {Verhamme}, O. and {Backs}, F. and {Van der Sijpt}, C. and {Sundqvist}, J.~O. and {Sander}, A.~A.~C.},
        title = "{2D unified atmosphere and wind simulations for a grid of O-type stars}",
      journal = {\aap},
         year = 2025,
        month = dec,
       volume = {704},
          eid = {A121},
        pages = {A121},
          doi = {10.1051/0004-6361/202556825},
archivePrefix = {arXiv},
       eprint = {2511.04542},
 primaryClass = {astro-ph.SR},
       adsurl = {https://ui.adsabs.harvard.edu/abs/2025A&A...704A.121M}
}

@ARTICLE{matheus_2025,
       author = {{Bernini-Peron}, M. and {Sander}, A.~A.~C. and {Ramachandran}, V. and {Oskinova}, L.~M. and {Vink}, J.~S. and {Verhamme}, O. and {Najarro}, F. and {Josiek}, J. and {Brands}, S.~A. and {Crowther}, P.~A. and {G{\'o}mez-Gonz{\'a}lez}, V.~M.~A. and {Gormaz-Matamala}, A.~C. and {Hawcroft}, C. and {Kuiper}, R. and {Mahy}, L. and {Marcolino}, W.~L.~F. and {Martins}, L.~P. and {Mehner}, A. and {Parsons}, T.~N. and {Pauli}, D. and {Shenar}, T. and {Schootemeijer}, A. and {Todt}, H. and {van Loon}, J. Th. and {XShootU Collaboration}},
        title = "{X-Shooting ULLYSES: Massive stars at low metallicity: VII. Stellar and wind properties of B supergiants in the Small Magellanic Cloud}",
      journal = {\aap},
         year = 2024,
        month = dec,
       volume = {692},
          eid = {A89},
        pages = {A89},
          doi = {10.1051/0004-6361/202450475},
archivePrefix = {arXiv},
       eprint = {2407.14216},
 primaryClass = {astro-ph.SR},
       adsurl = {https://ui.adsabs.harvard.edu/abs/2024A&A...692A..89B}
}

@ARTICLE{puls_2020,
       author = {{Puls}, J. and {Najarro}, F. and {Sundqvist}, J.~O. and {Sen}, K.},
        title = "{Atmospheric NLTE models for the spectroscopic analysis of blue stars with winds. V. Complete comoving frame transfer, and updated modeling of X-ray emission}",
      journal = {\aap},
         year = 2020,
        month = oct,
       volume = {642},
          eid = {A172},
        pages = {A172},
          doi = {10.1051/0004-6361/202038464},
archivePrefix = {arXiv},
       eprint = {2011.02310},
 primaryClass = {astro-ph.SR},
       adsurl = {https://ui.adsabs.harvard.edu/abs/2020A&A...642A.172P}
}

@ARTICLE{levermore_pomraning_1981,
       author = {{Levermore}, C.~D. and {Pomraning}, G.~C.},
        title = "{A flux-limited diffusion theory}",
      journal = {\apj},
         year = 1981,
        month = aug,
       volume = {248},
        pages = {321-334},
          doi = {10.1086/159157},
       adsurl = {https://ui.adsabs.harvard.edu/abs/1981ApJ...248..321L}
}

@ARTICLE{asif_2025,
       author = {{ud-Doula}, A. and {Sundqvist}, J.~O. and {Narechania}, N. and {Debnath}, D. and {Moens}, N. and {Keppens}, R.},
        title = "{Suppression of photospheric velocity fluctuations in strongly magnetic O stars in radiation-magnetohydrodynamic simulations}",
      journal = {\aap},
         year = 2025,
        month = jan,
       volume = {693},
          eid = {A224},
        pages = {A224},
          doi = {10.1051/0004-6361/202452375},
archivePrefix = {arXiv},
       eprint = {2412.10825},
 primaryClass = {astro-ph.SR},
       adsurl = {https://ui.adsabs.harvard.edu/abs/2025A&A...693A.224U}
}

@ARTICLE{pieter_2026,
       author = {{Schillemans}, P. and {Sundqvist}, J.~O. and {Debnath}, D. and {Delbroek}, L. and {Moens}, N. and {Van der Sijpt}, C.},
        title = "{Simulations of massive star atmospheres and winds during giant eruptive and quiescent luminous blue variable phases}",
      journal = {\aap},
         year = 2026,
        month = sep,
       volume = {713},
          eid = {A127},
        pages = {A127},
          doi = {10.1051/0004-6361/202660010},
archivePrefix = {arXiv},
       eprint = {2603.22168},
 primaryClass = {astro-ph.SR},
       adsurl = {https://ui.adsabs.harvard.edu/abs/2026A&A...713A.127S}
}

@ARTICLE{Debnath_2026,
       author = {{Debnath}, D. and {Sundqvist}, J.~O. and {Moens}, N. and {Poniatowski}, L.~G. and {Van der Sijpt}, C. and {Sander}, A.~A.~C.},
        title = "{Two-dimensional, time-dependent approximate NLTE unified model atmospheres with winds for hot, massive stars}",
      journal = {\aap},
         year = 2026,
        month = oct,
       volume = {714},
          eid = {A418},
        pages = {A418},
          doi = {10.1051/0004-6361/202660303},
archivePrefix = {arXiv},
       eprint = {2604.07317},
 primaryClass = {astro-ph.SR},
       adsurl = {https://ui.adsabs.harvard.edu/abs/2026A&A...714A.418D}
}

@ARTICLE{friend_castor_83,
       author = {{Friend}, D.~B. and {Castor}, J.~I.},
        title = "{Stellar winds driven by multiline scattering}",
      journal = {\apj},
         year = 1983,
        month = sep,
       volume = {272},
        pages = {259-272},
          doi = {10.1086/161289},
       adsurl = {https://ui.adsabs.harvard.edu/abs/1983ApJ...272..259F}
}

@ARTICLE{langer_2012,
       author = {{Langer}, N.},
        title = "{Presupernova Evolution of Massive Single and Binary Stars}",
      journal = {\araa},
         year = 2012,
        month = sep,
       volume = {50},
        pages = {107-164},
          doi = {10.1146/annurev-astro-081811-125534},
archivePrefix = {arXiv},
       eprint = {1206.5443},
 primaryClass = {astro-ph.SR},
       adsurl = {https://ui.adsabs.harvard.edu/abs/2012ARA&A..50..107L}
}

@ARTICLE{bjorklund_2023,
       author = {{Bj{\"o}rklund}, R. and {Sundqvist}, J.~O. and {Singh}, S.~M. and {Puls}, J. and {Najarro}, F.},
        title = "{New predictions for radiation-driven, steady-state mass-loss and wind-momentum from hot, massive stars. III. Updated mass-loss rates for stellar evolution}",
      journal = {\aap},
         year = 2023,
        month = aug,
       volume = {676},
          eid = {A109},
        pages = {A109},
          doi = {10.1051/0004-6361/202141948},
archivePrefix = {arXiv},
       eprint = {2203.08218},
 primaryClass = {astro-ph.SR},
       adsurl = {https://ui.adsabs.harvard.edu/abs/2023A&A...676A.109B}
}

@ARTICLE{puls_2008,
       author = {{Puls}, Joachim and {Vink}, Jorick S. and {Najarro}, Francisco},
        title = "{Mass loss from hot massive stars}",
      journal = {\aapr},
         year = 2008,
        month = dec,
       volume = {16},
       number = {3-4},
        pages = {209-325},
          doi = {10.1007/s00159-008-0015-8},
archivePrefix = {arXiv},
       eprint = {0811.0487},
 primaryClass = {astro-ph},
       adsurl = {https://ui.adsabs.harvard.edu/abs/2008A&ARv..16..209P}
}

@ARTICLE{kee_2021,
       author = {{Kee}, N.~D. and {Sundqvist}, J.~O. and {Decin}, L. and {de Koter}, A. and {Sana}, H.},
        title = "{Analytic, dust-independent mass-loss rates for red supergiant winds initiated by turbulent pressure}",
      journal = {\aap},
         year = 2021,
        month = feb,
       volume = {646},
          eid = {A180},
        pages = {A180},
          doi = {10.1051/0004-6361/202039224},
archivePrefix = {arXiv},
       eprint = {2101.03070},
 primaryClass = {astro-ph.SR},
       adsurl = {https://ui.adsabs.harvard.edu/abs/2021A&A...646A.180K}
}

@ARTICLE{sander_2020,
       author = {{Sander}, Andreas A.~C. and {Vink}, J.~S. and {Hamann}, W.-R.},
        title = "{Driving classical Wolf-Rayet winds: a {\ensuremath{\Gamma}}- and Z-dependent mass-loss}",
      journal = {\mnras},
         year = 2020,
        month = jan,
       volume = {491},
       number = {3},
        pages = {4406-4425},
          doi = {10.1093/mnras/stz3064},
archivePrefix = {arXiv},
       eprint = {1910.12886},
 primaryClass = {astro-ph.SR},
       adsurl = {https://ui.adsabs.harvard.edu/abs/2020MNRAS.491.4406S}
}

@ARTICLE{grafener_2005,
       author = {{Gr{\"a}fener}, G. and {Hamann}, W.-R.},
        title = "{Hydrodynamic model atmospheres for WR stars. Self-consistent modeling of a WC star wind}",
      journal = {\aap},
         year = 2005,
        month = mar,
       volume = {432},
       number = {2},
        pages = {633-645},
          doi = {10.1051/0004-6361:20041732},
archivePrefix = {arXiv},
       eprint = {astro-ph/0410697},
 primaryClass = {astro-ph},
       adsurl = {https://ui.adsabs.harvard.edu/abs/2005A&A...432..633G}
}

@ARTICLE{verhamme_2026,
       author = {{Verhamme}, O. and {Sundqvist}, J.~O. and {de Koter}, A. and {Sana}, H. and {Backs}, F. and {Brands}, S.~A. and {Debnath}, D. and {Moens}, N. and {Schillemans}, P. and {Van der Sijpt}, C. and {Berlanas}, S.~R. and {Bernini-Peron}, M. and {Crowther}, P.~A. and {Gormaz-Matamala}, A.~C. and {Kuiper}, R. and {Hawcroft}, C. and {Najarro}, F. and {Pauli}, D. and {Sander}, A.~A.~C. and {van Loon}, J. Th. and {Vink}, J.~S. and {Todt}, H. and {Tramper}, F. and {Xshootu collaboration}},
        title = "{X-shooting ULLYSES: Massive stars at low metallicity: XV. On the metallicity dependence of B-supergiant mass-loss rates}",
      journal = {\aap},
         year = 2026,
        month = jun,
       volume = {710},
          eid = {A166},
        pages = {A166},
          doi = {10.1051/0004-6361/202558274},
archivePrefix = {arXiv},
       eprint = {2604.21452},
 primaryClass = {astro-ph.SR},
       adsurl = {https://ui.adsabs.harvard.edu/abs/2026A&A...710A.166V}
}

@ARTICLE{markova_2008,
       author = {{Markova}, N. and {Puls}, J.},
        title = "{Bright OB stars in the Galaxy. IV. Stellar and wind parameters of early to late B supergiants}",
      journal = {\aap},
         year = 2008,
        month = feb,
       volume = {478},
       number = {3},
        pages = {823-842},
          doi = {10.1051/0004-6361:20077919},
archivePrefix = {arXiv},
       eprint = {0711.1110},
 primaryClass = {astro-ph},
       adsurl = {https://ui.adsabs.harvard.edu/abs/2008A&A...478..823M}
}

@BOOK{mihalas_1978,
       author = {{Mihalas}, Dimitri},
        title = "{Stellar atmospheres}",
         year = 1978,
       adsurl = {https://ui.adsabs.harvard.edu/abs/1978stat.book.....M}
}

@ARTICLE{gonzalez_2025,
       author = {{Gonz{\'a}lez-Tor{\`a}}, G. and {Sander}, A.~A.~C. and {Sundqvist}, J.~O. and {Debnath}, D. and {Delbroek}, L. and {Josiek}, J. and {Lefever}, R.~R. and {Moens}, N. and {Van der Sijpt}, C. and {Verhamme}, O.},
        title = "{Improving 1D stellar atmosphere models with insights from multi-dimensional simulations: I. 1D versus 2D stratifications and spectral comparisons for O stars}",
      journal = {\aap},
         year = 2025,
        month = feb,
       volume = {694},
          eid = {A269},
        pages = {A269},
          doi = {10.1051/0004-6361/202452241},
archivePrefix = {arXiv},
       eprint = {2501.14511},
 primaryClass = {astro-ph.SR},
       adsurl = {https://ui.adsabs.harvard.edu/abs/2025A&A...694A.269G}
}

@ARTICLE{kriticka_2017,
       author = {{Krti{\v{c}}ka}, J. and {Kub{\'a}t}, J.},
        title = "{Comoving frame models of hot star winds. II. Reduction of O star wind mass-loss rates in global models}",
      journal = {\aap},
         year = 2017,
        month = oct,
       volume = {606},
          eid = {A31},
        pages = {A31},
          doi = {10.1051/0004-6361/201730723},
archivePrefix = {arXiv},
       eprint = {1706.06194},
 primaryClass = {astro-ph.SR},
       adsurl = {https://ui.adsabs.harvard.edu/abs/2017A&A...606A..31K}
}

@ARTICLE{lamers_1995,
       author = {{Lamers}, Henny J.~G.~L.~M. and {Snow}, Theodore P. and {Lindholm}, Douglas M.},
        title = "{Terminal Velocities and the Bistability of Stellar Winds}",
      journal = {\apj},
         year = 1995,
        month = dec,
       volume = {455},
        pages = {269},
          doi = {10.1086/176575},
       adsurl = {https://ui.adsabs.harvard.edu/abs/1995ApJ...455..269L}
}

@ARTICLE{pauldrach_puls_1990,
       author = {{Pauldrach}, A.~W.~A. and {Puls}, J.},
        title = "{Radiation-driven winds of hot luminous stars. VIII. The bistable wind of the luminous blue variable P Cygni (B1 Ia+).}",
      journal = {\aap},
         year = 1990,
        month = oct,
       volume = {237},
        pages = {409},
       adsurl = {https://ui.adsabs.harvard.edu/abs/1990A&A...237..409P}
}

@ARTICLE{crowther_2006,
       author = {{Crowther}, P.~A. and {Lennon}, D.~J. and {Walborn}, N.~R.},
        title = "{Physical parameters and wind properties of galactic early B supergiants}",
      journal = {\aap},
         year = 2006,
        month = jan,
       volume = {446},
       number = {1},
        pages = {279-293},
          doi = {10.1051/0004-6361:20053685},
archivePrefix = {arXiv},
       eprint = {astro-ph/0509436},
 primaryClass = {astro-ph},
       adsurl = {https://ui.adsabs.harvard.edu/abs/2006A&A...446..279C}
}

@ARTICLE{de_Burgos_2024,
       author = {{de Burgos}, A. and {Sim{\'o}n-D{\'\i}az}, S. and {Urbaneja}, M.~A. and {Puls}, J.},
        title = "{The IACOB project. X. Large-scale quantitative spectroscopic analysis of Galactic luminous blue stars}",
      journal = {\aap},
         year = 2024,
        month = jul,
       volume = {687},
          eid = {A228},
        pages = {A228},
          doi = {10.1051/0004-6361/202348808},
archivePrefix = {arXiv},
       eprint = {2312.00241},
 primaryClass = {astro-ph.SR},
       adsurl = {https://ui.adsabs.harvard.edu/abs/2024A&A...687A.228D}
}

@ARTICLE{kriticka_2021,
       author = {{Krti{\v{c}}ka}, J. and {Kub{\'a}t}, J. and {Krti{\v{c}}kov{\'a}}, I.},
        title = "{New mass-loss rates of B supergiants from global wind models}",
      journal = {\aap},
         year = 2021,
        month = mar,
       volume = {647},
          eid = {A28},
        pages = {A28},
          doi = {10.1051/0004-6361/202039900},
archivePrefix = {arXiv},
       eprint = {2101.04973},
 primaryClass = {astro-ph.SR},
       adsurl = {https://ui.adsabs.harvard.edu/abs/2021A&A...647A..28K}
}

@ARTICLE{haucke_2018,
       author = {{Haucke}, M. and {Cidale}, L.~S. and {Venero}, R.~O.~J. and {Cur{\'e}}, M. and {Kraus}, M. and {Kanaan}, S. and {Arcos}, C.},
        title = "{Wind properties of variable B supergiants. Evidence of pulsations connected with mass-loss episodes}",
      journal = {\aap},
         year = 2018,
        month = jun,
       volume = {614},
          eid = {A91},
        pages = {A91},
          doi = {10.1051/0004-6361/201731678},
archivePrefix = {arXiv},
       eprint = {1902.01341},
 primaryClass = {astro-ph.SR},
       adsurl = {https://ui.adsabs.harvard.edu/abs/2018A&A...614A..91H}
}

@ARTICLE{Cure04,
       author = {{Cur{\'e}}, Michel},
        title = "{The Influence of Rotation in Radiation-driven Wind from Hot Stars: New Solutions and Disk Formation in Be Stars}",
      journal = {\apj},
         year = 2004,
        month = oct,
       volume = {614},
       number = {2},
        pages = {929-941},
          doi = {10.1086/423776},
archivePrefix = {arXiv},
       eprint = {astro-ph/0406490},
 primaryClass = {astro-ph},
       adsurl = {https://ui.adsabs.harvard.edu/abs/2004ApJ...614..929C}
}

@misc{verhamme_thesis_2025,
author = {Verhamme, Olivier},
language = {eng},
title = {Spectroscopy of massive stars at low metallicity: Studying the atmospheres and outflows of B-stars},
year = {2025},
}

@misc{moens_thesis_2023,
author = {Moens, Nicolas},
language = {eng},
title = {Radiation hydrodynamics of massive star outflows},
year = {2023},
}

@article{venero_2024,
    author = {Venero, R O J and Curé, M and Puls, J and Cidale, L S and Haucke, M and Araya, I and Gormaz-Matamala, A and Arcos, C},
    title = {The wind of rotating B supergiants – II. The δ-slow hydrodynamic regime},
    journal = {Monthly Notices of the Royal Astronomical Society},
    volume = {527},
    number = {1},
    pages = {93-111},
    year = {2024},
    month = {01},
    issn = {0035-8711},
    doi = {10.1093/mnras/stad3030},
    url = {https://doi.org/10.1093/mnras/stad3030},
    eprint = {https://academic.oup.com/mnras/article-pdf/527/1/93/53003258/stad3030.pdf},
}

@ARTICLE{cure_11,
       author = {{Cur{\'e}}, M. and {Cidale}, L. and {Granada}, A.},
        title = "{Slow Radiation-driven Wind Solutions of A-type Supergiants}",
      journal = {\apj},
         year = 2011,
        month = aug,
       volume = {737},
       number = {1},
          eid = {18},
        pages = {18},
          doi = {10.1088/0004-637X/737/1/18},
archivePrefix = {arXiv},
       eprint = {1105.5576},
 primaryClass = {astro-ph.SR},
       adsurl = {https://ui.adsabs.harvard.edu/abs/2011ApJ...737...18C}
}

@ARTICLE{venero_2016,
       author = {{Venero}, R.~O.~J. and {Cur{\'e}}, M. and {Cidale}, L.~S. and {Araya}, I.},
        title = "{The Wind of Rotating B Supergiants. I. Domains of Slow and Fast Solution Regimes}",
      journal = {\apj},
         year = 2016,
        month = may,
       volume = {822},
       number = {1},
          eid = {28},
        pages = {28},
          doi = {10.3847/0004-637X/822/1/28},
       adsurl = {https://ui.adsabs.harvard.edu/abs/2016ApJ...822...28V}
}

@ARTICLE{vink_2000,
       author = {{Vink}, J.~S. and {de Koter}, A. and {Lamers}, H.~J.~G.~L.~M.},
        title = "{New theoretical mass-loss rates of O and B stars}",
      journal = {\aap},
         year = 2000,
        month = oct,
       volume = {362},
        pages = {295-309},
          doi = {10.48550/arXiv.astro-ph/0008183},
archivePrefix = {arXiv},
       eprint = {astro-ph/0008183},
 primaryClass = {astro-ph},
       adsurl = {https://ui.adsabs.harvard.edu/abs/2000A&A...362..295V}
}

@ARTICLE{Kudritzki_1999,
       author = {{Kudritzki}, R.~P. and {Puls}, J. and {Lennon}, D.~J. and {Venn}, K.~A. and {Reetz}, J. and {Najarro}, F. and {McCarthy}, J.~K. and {Herrero}, A.},
        title = "{The wind momentum-luminosity relationship of galactic A- and B-supergiants}",
      journal = {\aap},
         year = 1999,
        month = oct,
       volume = {350},
        pages = {970-984},
          doi = {10.48550/arXiv.astro-ph/9910449},
archivePrefix = {arXiv},
       eprint = {astro-ph/9910449},
 primaryClass = {astro-ph},
       adsurl = {https://ui.adsabs.harvard.edu/abs/1999A&A...350..970K}
}

@ARTICLE{lefever_2007,
       author = {{Lefever}, K. and {Puls}, J. and {Aerts}, C.},
        title = "{Statistical properties of a sample of periodically variable B-type supergiants. Evidence for opacity-driven gravity-mode oscillations}",
      journal = {\aap},
         year = 2007,
        month = mar,
       volume = {463},
       number = {3},
        pages = {1093-1109},
          doi = {10.1051/0004-6361:20066038},
archivePrefix = {arXiv},
       eprint = {astro-ph/0611484},
 primaryClass = {astro-ph},
       adsurl = {https://ui.adsabs.harvard.edu/abs/2007A&A...463.1093L}
}

@ARTICLE{Searle_2008,
       author = {{Searle}, S.~C. and {Prinja}, R.~K. and {Massa}, D. and {Ryans}, R.},
        title = "{Quantitative studies of the optical and UV spectra of Galactic early B supergiants. I. Fundamental parameters}",
      journal = {\aap},
         year = 2008,
        month = apr,
       volume = {481},
       number = {3},
        pages = {777-797},
          doi = {10.1051/0004-6361:20077125},
archivePrefix = {arXiv},
       eprint = {0801.4289},
 primaryClass = {astro-ph},
       adsurl = {https://ui.adsabs.harvard.edu/abs/2008A&A...481..777S}
}

@ARTICLE{Vink_2026,
       author = {{Vink}, Jorick S. and {Sabhahit}, Gautham N. and {Sander}, Andreas A.~C.},
        title = "{The robustness of bi-stability jump predictions}",
      journal = {\aap},
         year = 2026,
        month = jul,
       volume = {711},
          eid = {L4},
        pages = {L4},
          doi = {10.1051/0004-6361/202659949},
archivePrefix = {arXiv},
       eprint = {2606.17130},
 primaryClass = {astro-ph.SR},
       adsurl = {https://ui.adsabs.harvard.edu/abs/2026A&A...711L...4V}
}

@ARTICLE{Bernini-Peron_2026,
       author = {{Bernini-Peron}, M. and {Sander}, A.~A.~C. and {Sabhahit}, G.~N. and {Najarro}, F. and {Ramachandran}, V. and {Vink}, J.~S.},
        title = "{The winds of OBA hypergiants and luminous blue variables: Dynamically consistent atmosphere models reveal multiple wind regimes}",
      journal = {\aap},
         year = 2026,
        month = apr,
       volume = {708},
          eid = {A206},
        pages = {A206},
          doi = {10.1051/0004-6361/202558762},
archivePrefix = {arXiv},
       eprint = {2601.03072},
 primaryClass = {astro-ph.SR},
       adsurl = {https://ui.adsabs.harvard.edu/abs/2026A&A...708A.206B}
}

@misc{Ma_2026,
      title={3D Radiation Hydrodynamics of Blue Supergiant Envelopes}, 
      author={Linhao Ma and Lars Bildsten and Yan-Fei Jiang},
      year={2026},
      eprint={2609.28656},
      archivePrefix={arXiv},
      primaryClass={astro-ph.SR},
      url={https://arxiv.org/abs/2609.28656}, 
}

@ARTICLE{Bowman2019,
       author = {{Bowman}, Dominic M. and {Burssens}, Siemen and {Pedersen}, May G. and {Johnston}, Cole and {Aerts}, Conny and {Buysschaert}, Bram and {Michielsen}, Mathias and {Tkachenko}, Andrew and {Rogers}, Tamara M. and {Edelmann}, Philipp V.~F. and {Ratnasingam}, Rathish P. and {Sim{\'o}n-D{\'\i}az}, Sergio and {Castro}, Norberto and {Moravveji}, Ehsan and {Pope}, Benjamin J.~S. and {White}, Timothy R. and {De Cat}, Peter},
        title = "{Low-frequency gravity waves in blue supergiants revealed by high-precision space photometry}",
      journal = {Nature Astronomy},
         year = 2019,
        month = may,
       volume = {3},
        pages = {760-765},
          doi = {10.1038/s41550-019-0768-1},
archivePrefix = {arXiv},
       eprint = {1905.02120},
 primaryClass = {astro-ph.SR},
       adsurl = {https://ui.adsabs.harvard.edu/abs/2019NatAs...3..760B}
}

@ARTICLE{Ma_2024,
       author = {{Ma}, Linhao and {Johnston}, Cole and {Bellinger}, Earl Patrick and {de Mink}, Selma E.},
        title = "{Variability of Blue Supergiants in the LMC with TESS}",
      journal = {\apj},
         year = 2024,
        month = may,
       volume = {966},
       number = {2},
          eid = {196},
        pages = {196},
          doi = {10.3847/1538-4357/ad38bc},
archivePrefix = {arXiv},
       eprint = {2310.19546},
 primaryClass = {astro-ph.SR},
       adsurl = {https://ui.adsabs.harvard.edu/abs/2024ApJ...966..196M}
}

@ARTICLE{Kourniotis_2025,
       author = {{Kourniotis}, Michalis and {Cidale}, Lydia S. and {Kraus}, Michaela and {Ruiz Diaz}, Matias A. and {Alberici Adam}, Aldana},
        title = "{Variability of Galactic blue supergiants observed with TESS}",
      journal = {\aap},
         year = 2025,
        month = may,
       volume = {697},
          eid = {A152},
        pages = {A152},
          doi = {10.1051/0004-6361/202452360},
archivePrefix = {arXiv},
       eprint = {2503.20860},
 primaryClass = {astro-ph.SR},
       adsurl = {https://ui.adsabs.harvard.edu/abs/2025A&A...697A.152K}
}

@ARTICLE{Bowman2019b,
       author = {{Bowman}, D.~M. and {Aerts}, C. and {Johnston}, C. and {Pedersen}, M.~G. and {Rogers}, T.~M. and {Edelmann}, P.~V.~F. and {Sim{\'o}n-D{\'\i}az}, S. and {Van Reeth}, T. and {Buysschaert}, B. and {Tkachenko}, A. and {Triana}, S.~A.},
        title = "{Photometric detection of internal gravity waves in upper main-sequence stars. I. Methodology and application to CoRoT targets}",
      journal = {\aap},
         year = 2019,
        month = jan,
       volume = {621},
          eid = {A135},
        pages = {A135},
          doi = {10.1051/0004-6361/201833662},
archivePrefix = {arXiv},
       eprint = {1811.08023},
 primaryClass = {astro-ph.SR},
       adsurl = {https://ui.adsabs.harvard.edu/abs/2019A&A...621A.135B}
}

@ARTICLE{Bowman2020,
       author = {{Bowman}, D.~M. and {Burssens}, S. and {Sim{\'o}n-D{\'\i}az}, S. and {Edelmann}, P.~V.~F. and {Rogers}, T.~M. and {Horst}, L. and {R{\"o}pke}, F.~K. and {Aerts}, C.},
        title = "{Photometric detection of internal gravity waves in upper main-sequence stars. II. Combined TESS photometry and high-resolution spectroscopy}",
      journal = {\aap},
         year = 2020,
        month = aug,
       volume = {640},
          eid = {A36},
        pages = {A36},
          doi = {10.1051/0004-6361/202038224},
archivePrefix = {arXiv},
       eprint = {2006.03012},
 primaryClass = {astro-ph.SR},
       adsurl = {https://ui.adsabs.harvard.edu/abs/2020A&A...640A..36B}
}

@article{Ryans_2002,
    author = {Ryans, R. S. I. and Dufton, P. L. and Rolleston, W. R. J. and Lennon, D. J. and Keenan, F. P. and Smoker, J. V. and Lambert, D. L.},
    title = {Macroturbulent and rotational broadening in the spectra of B-type supergiants},
    journal = {Monthly Notices of the Royal Astronomical Society},
    volume = {336},
    number = {2},
    pages = {577-586},
    year = {2002},
    month = {10},
    issn = {0035-8711},
    doi = {10.1046/j.1365-8711.2002.05780.x},
    url = {https://doi.org/10.1046/j.1365-8711.2002.05780.x},
    eprint = {https://academic.oup.com/mnras/article-pdf/336/2/577/19571909/336-2-577.pdf},
}

@ARTICLE{nadya_2024,
       author = {{Serebriakova}, Nadya and {Tkachenko}, Andrew and {Aerts}, Conny},
        title = "{The ESO UVES/FEROS Large Programs of TESS OB pulsators: II. The physical origin of macroturbulence}",
      journal = {\aap},
         year = 2024,
        month = dec,
       volume = {692},
          eid = {A245},
        pages = {A245},
          doi = {10.1051/0004-6361/202451581},
archivePrefix = {arXiv},
       eprint = {2408.15888},
 primaryClass = {astro-ph.SR},
       adsurl = {https://ui.adsabs.harvard.edu/abs/2024A&A...692A.245S}
}

@ARTICLE{Ricker_2015,
       author = {{Ricker}, George R. and {Winn}, Joshua N. and {Vanderspek}, Roland and {Latham}, David W. and {Bakos}, G{\'a}sp{\'a}r {\'A}. and {Bean}, Jacob L. and {Berta-Thompson}, Zachory K. and {Brown}, Timothy M. and {Buchhave}, Lars and {Butler}, Nathaniel R. and {Butler}, R. Paul and {Chaplin}, William J. and {Charbonneau}, David and {Christensen-Dalsgaard}, J{\o}rgen and {Clampin}, Mark and {Deming}, Drake and {Doty}, John and {De Lee}, Nathan and {Dressing}, Courtney and {Dunham}, Edward W. and {Endl}, Michael and {Fressin}, Francois and {Ge}, Jian and {Henning}, Thomas and {Holman}, Matthew J. and {Howard}, Andrew W. and {Ida}, Shigeru and {Jenkins}, Jon M. and {Jernigan}, Garrett and {Johnson}, John Asher and {Kaltenegger}, Lisa and {Kawai}, Nobuyuki and {Kjeldsen}, Hans and {Laughlin}, Gregory and {Levine}, Alan M. and {Lin}, Douglas and {Lissauer}, Jack J. and {MacQueen}, Phillip and {Marcy}, Geoffrey and {McCullough}, Peter R. and {Morton}, Timothy D. and {Narita}, Norio and {Paegert}, Martin and {Palle}, Enric and {Pepe}, Francesco and {Pepper}, Joshua and {Quirrenbach}, Andreas and {Rinehart}, Stephen A. and {Sasselov}, Dimitar and {Sato}, Bun'ei and {Seager}, Sara and {Sozzetti}, Alessandro and {Stassun}, Keivan G. and {Sullivan}, Peter and {Szentgyorgyi}, Andrew and {Torres}, Guillermo and {Udry}, Stephane and {Villasenor}, Joel},
        title = "{Transiting Exoplanet Survey Satellite (TESS)}",
      journal = {Journal of Astronomical Telescopes, Instruments, and Systems},
         year = 2015,
        month = jan,
       volume = {1},
          eid = {014003},
        pages = {014003},
          doi = {10.1117/1.JATIS.1.1.014003},
archivePrefix = {arXiv},
       eprint = {1406.0151},
 primaryClass = {astro-ph.EP},
       adsurl = {https://ui.adsabs.harvard.edu/abs/2015JATIS...1a4003R}
}

@ARTICLE{Grassitelli_2015,
       author = {{Grassitelli}, L. and {Fossati}, L. and {Langer}, N. and {Miglio}, A. and {Istrate}, A.~G. and {Sanyal}, D.},
        title = "{Relating turbulent pressure and macroturbulence across the HR diagram with a possible link to {\ensuremath{\gamma}} Doradus stars}",
      journal = {\aap},
         year = 2015,
        month = dec,
       volume = {584},
          eid = {L2},
        pages = {L2},
          doi = {10.1051/0004-6361/201527289},
archivePrefix = {arXiv},
       eprint = {1511.01487},
 primaryClass = {astro-ph.SR},
       adsurl = {https://ui.adsabs.harvard.edu/abs/2015A&A...584L...2G}
}

@ARTICLE{Vink_1999,
       author = {{Vink}, J.~S. and {de Koter}, A. and {Lamers}, H.~J.~G.~L.~M.},
        title = "{On the nature of the bi-stability jump in the winds of early-type supergiants}",
      journal = {\aap},
         year = 1999,
        month = oct,
       volume = {350},
        pages = {181-196},
          doi = {10.48550/arXiv.astro-ph/9908196},
archivePrefix = {arXiv},
       eprint = {astro-ph/9908196},
 primaryClass = {astro-ph},
       adsurl = {https://ui.adsabs.harvard.edu/abs/1999A&A...350..181V}
}

@ARTICLE{Prinja_2002,
       author = {{Prinja}, R.~K. and {Massa}, D. and {Fullerton}, A.~W.},
        title = "{Wind variability of B supergiants. IV. A survey of IUE time-series data of 11 B0 to B3 stars}",
      journal = {\aap},
         year = 2002,
        month = jun,
       volume = {388},
        pages = {587-608},
          doi = {10.1051/0004-6361:20020503},
       adsurl = {https://ui.adsabs.harvard.edu/abs/2002A&A...388..587P}
}

@ARTICLE{Lomb_1976,
       author = {{Lomb}, N.~R.},
        title = "{Least-Squares Frequency Analysis of Unequally Spaced Data}",
      journal = {\apss},
         year = 1976,
        month = feb,
       volume = {39},
       number = {2},
        pages = {447-462},
          doi = {10.1007/BF00648343},
       adsurl = {https://ui.adsabs.harvard.edu/abs/1976Ap&SS..39..447L}
}

@ARTICLE{Scargle_1982,
       author = {{Scargle}, J.~D.},
        title = "{Studies in astronomical time series analysis. II. Statistical aspects of spectral analysis of unevenly spaced data.}",
      journal = {\apj},
         year = 1982,
        month = dec,
       volume = {263},
        pages = {835-853},
          doi = {10.1086/160554},
       adsurl = {https://ui.adsabs.harvard.edu/abs/1982ApJ...263..835S}
}

@article{Pathak_2026,
doi = {10.3847/1538-4357/ae4487},
url = {https://doi.org/10.3847/1538-4357/ae4487},
year = {2026},
month = {mar},
publisher = {The American Astronomical Society},
volume = {1000},
number = {1},
pages = {89},
author = {Pathak, Praneet and Blouin, Simon and Herwig, Falk and Woodward, Paul R.},
title = {3D Hydrodynamic Simulations of Massive Main-sequence Stars. IV. Internal Gravity Wave Matter for Stochastic Low-frequency Variability},
journal = {The Astrophysical Journal}
}

@ARTICLE{ankur_2025,
       author = {{Kalita}, A.~J. and {Bowman}, D.~M. and {Abdul-Masih}, M. and {Sim{\'o}n-D{\'\i}az}, S.},
        title = "{Large-scale variability in macroturbulence driven by pulsations in the rapidly rotating massive star {\ensuremath{\zeta}} Oph from high-cadence ESPRESSO spectroscopy and TESS photometry}",
      journal = {\aap},
         year = 2025,
        month = nov,
       volume = {703},
          eid = {A2},
        pages = {A2},
          doi = {10.1051/0004-6361/202555936},
archivePrefix = {arXiv},
       eprint = {2509.09635},
 primaryClass = {astro-ph.SR},
       adsurl = {https://ui.adsabs.harvard.edu/abs/2025A&A...703A...2K}
}

\end{document}